\documentclass{aa}  
\usepackage{natbib}
\bibpunct{(}{)}{;}{a}{}{,} % to follow the A&A style
\usepackage{pbox}
\usepackage{url}
\usepackage{amssymb}

\usepackage{soul}
\usepackage[colorlinks=true,linkcolor=blue,citecolor=blue]{hyperref}

\usepackage{graphicx}
\usepackage{txfonts}

\begin{document}

   \title{Chromatic transit light curves of  disintegrating rocky planets} 
   \titlerunning{Chromatic transit light curves of disintegrating rocky planets}

   \author{A. R. Ridden-Harper \inst{1}
          \and C. U. Keller \inst{1}          
          \and M. Min \inst{2} 
          \and R. van Lieshout \inst{3} 
          \and I. A. G. Snellen \inst{1}}

   \institute{Leiden Observatory, Leiden University, Niels Bohrweg 2, 2333 CA Leiden, The Netherlands
                  \\
              \email{arh@strw.leidenuniv.nl}              
             \and 
             Netherlands Institute for Space Research (SRON), Sorbonnelaan 2, 3584 CA Utrecht, The Netherlands          
              \and
             Institute of Astronomy, University of Cambridge, Madingley Rd, Cambridge CB3 0HA, UK}

   \date{}

% \abstract{}{}{}{}{} 
% 5 {} token are mandatory

  \abstract
   {Kepler observations have revealed a class of short period exoplanets, of which Kepler-1520 b is the prototype, which have comet-like dust tails thought to be the
   result of small, rocky planets losing mass. The shape and chromaticity of the transits constrain the properties of the dust particles originating from the planet's surface, offering a unique opportunity to probe the composition and geophysics of rocky exoplanets.}
   {We aim to approximate the average Kepler long-cadence light curve of Kepler-1520 b and investigate how the optical thickness  and transit cross-section of a general dust tail can affect the observed wavelength dependence and depth of transit light curves.}
   {We developed a new 3D model that ejects sublimating particles from the planet surface to build up a dust tail, assuming it to be optically thin, and used 3D radiative transfer computations that fully treat scattering using the distribution of hollow spheres (DHS) method, to generate transit light curves between 0.45 and 2.5 $\mu$m.}
   {We show that the transit depth is wavelength independent for optically thick tails, potentially explaining why only some observations indicate a wavelength dependence.  From the 3D nature of our simulated tails, we show that their transit cross-sections are related to the component of particle ejection velocity perpendicular to the planet’s orbital plane and use this to derive a minimum ejection velocity of 1.2  kms$^{-1}$.  To fit the average transit depth of Kepler-1520 b of 0.87\%, we require a high dust mas-loss rate of 7 $-$ 80 M$_\oplus$ Gyr$^{-1}$ which  implies planet lifetimes that may be inconsistent with the observed sample.  Therefore, these mass-loss rates should be considered to be upper limits.}
   {}  


   \maketitle

%
%________________________________________________________________
\section{Introduction \label{sec:intro}}

Exoplanetary systems are found to exhibit a large diversity in system architecture, planet size, composition and temperature. An intriguing recent addition to this diversity is the class of close-in, rocky exoplanets that have large comet-like tails, consisting of dust particles that are thought to originate from the rocky planet as a result of the rocky planet losing mass.

The term comet-like tail was first used in an explanatory context by \cite{Vidal-Madjar2003} to describe their discovery of a stream of hydrogen atoms escaping from  the evaporating atmosphere of the hot-Jupiter type exoplanet HD209458 b and has subsequently been used to describe three other similar planets.  A similarity between the gas- and dust-tails is that they are both shaped by radiation pressure\footnote{Although, the tail of HD 189733 b has a large blue shift that is best explained by charge exchange interaction with the stellar wind. \citep{Vidal-Madjar2003,Ehrenreich2012,Rappaport2012,Bourrier2014,Ehrenreich2015,Holmstrom2008,Ekenback2010,Bourrier2013,Bourrier2013b}.}.

The transit light curves produced by dust tails from disintegrating rocky exoplanets are asymmetrical due to the extended tails decreasing in density along the tail, away from the planet.  They also feature forward-scattering peaks at ingress, and in some cases, egress  \citep[e.g.][]{Rappaport2012,Rappaport2014,Brogi2012b,vanLieshout2014,Sanchis-Ojeda2015,vanLieshout2016}.   To date, three such planets around main-sequence stars and one around a white dwarf have been discovered from Kepler light curves: Kepler-1520 b (also known as KIC 12557548 b) \citep{Rappaport2012}, KOI-2700 b \citep{Rappaport2014}, K2-22 b \citep{Sanchis-Ojeda2015} and WD 1145+017 \citep{Vanderburg2015}.  These planets all have orbital periods of less than one day and exhibit variable transit depths, with WD 1145+017 exhibiting transit depths of up to 40\%.  The dust in the tails of disintegrating rocky exoplanets originates from the outer parts of the planet.  Therefore, these objects present the exciting opportunity to observationally probe the outer composition of rocky exoplanets, which would be very valuable information for constraining models of their structure and geophysics.

Kepler-1520 b has been relatively well studied, and constraints on its mean particle size,  mass-loss rate (assuming an optically thin tail) and particle composition have been determined by fitting models to the Kepler light curves, and by searching for a wavelength dependence in the transit depth with spectrophotometric observations. The first constraints on the particle size and mass-loss rate for Kepler-1520 b were derived by \cite{Rappaport2012} in their discovery paper. Assuming an optically thin tail, they derived a mass loss rate of 1 M$_\oplus$Gyr$^{-1}$.  This was further refined by \cite{Perez-Becker2013} who show with improved models that for possibly porous grains with radius $>$0.1 $\mathrm{\mu}$m, the mass loss rate can have a lower value of $\gtrsim0.$1 M$_\oplus$Gyr$^{-1}$. \cite{Brogi2012b} develop a one-dimensional model of the dust tail with an exponentially decaying angular dust density away from the planet and derived a typical particle size of 0.1 $\mathrm{\mu}$m.  A complementary study carried out by  \cite{Budaj2013} modelled the dust tail as a complete or partial ring where the density varied as a power law or an exponential as a function of angular distance from the planet.   One of their main results is that the system was found to be best modelled with at least two components, one consisting of the transit core and the other producing the tail.  This is validated by \cite{vanWerkhoven2014} with their  implementation of a two-dimensional, two component model consisting of an exponential tail and an opaque core, which gave an improved fit to the Kepler short-cadence light curves.

Some interesting constraints have been applied to the composition of the dust particles in the tail of Kepler-1520 b.  \cite{vanLieshout2014} find the grains to be consistent with being composed of corundum (Al$_2$O$_3$) or iron-rich silicate materials.  This work is extended by \cite{vanLieshout2016} in which a self-consistent numerical model was developed to calculate the dynamics of the sublimating dust particles and generate synthetic light curves. They find that good fits to the observed light curves can be obtained with initial particle sizes between 0.2 and 5.6 $\mathrm{\mu}$m and mass-loss rates of 0.6 to 15.6 M$_\oplus$ Gyr$^{-1}$. Furthermore, they find the dust composition to be consistent with corundum (Al$_2$O$_3$) but not with several carbonaceous, silicate or iron compositions.

In addition to fitting the average Kepler light curves, information about the particle composition and size can be derived from spectrophotometric observations.  \cite{Croll2014} observe transits of Kepler-1520 b at 2.15 $\mathrm{\mu}$m, 0.53 $\mathrm{\mu}$m $-$ 0.77 $\mu$m and utilise the Kepler light curve at 0.6 $\mathrm{\mu}$m and found no wavelength dependence in transit depth.  They report that if the observed scattering was due to particles of a single size, the particles would have to be at least 0.5 $\mathrm{\mu}$m in radius. \cite{Felipe2013} observed three transits and one secondary eclipse of Kepler-1520 b with OSIRIS on the GTC and also found no evidence for a wavelength dependence in transit depth.

\cite{Schlawin2016} carried out a complementary search for a wavelength dependence in the transit depth of Kepler-1520 b to constrain the particle size. They observe eight transits with the SpeX spectrograph and the MORIS imager on the Infrared Telescope Facility, with a wavelength coverage of 0.6 $-$ 2.4 $\mu$m, and one night in H band (1.63 $\mu$m). They report a flat transmission spectrum, consistent with the particles being $\gtrsim$0.5 $\mathrm{\mu}$m for pyroxene and olivine or $\gtrsim$0.2 $\mathrm{\mu}$m for iron and corundum.

\cite{Bochinski2015} observe five transits with ULTRACAM on the 4.2 m William Herschel Telescope.  In contrast to the previously discussed results which indicate a wavelength independent transit depth, \cite{Bochinski2015} report a wavelength dependence to a confidence of 3.2$\sigma$.  These transit depths are consistent with absorption by interstellar medium (ISM) like material with grain sizes corresponding to the largest found in the ISM of 0.25 - 1 $\mu$m.

The exoplanet K2-22 b was discovered and characterised by \cite{Sanchis-Ojeda2015}.  They observe transits with several ground based 1m class telescopes and the Gran Telescopio Canarias (GTC).  Their observations reveal it to have highly variable transit depths from 0 to 1.3\%, variable  transit shapes, and on one occasion, a significant wavelength dependence.  They infer that the distribution of dust particle sizes (a) must be a non-steep  power-law, $dN/da \propto a^{-\Gamma}$ with $\Gamma \simeq 1 - 3$ with maximum sizes in the range of 0.4 $-$ 0.7 $\mu$m.  They also determine its tail to be leading (instead of trailing) the planet.  The leading tail requires the dust to be transported to a distance of about twice the planetary radius towards the host star where it effectively overflows the planet's Roche lobe and goes into a faster orbit than that of the planet, allowing it to move in front of the planet.  This can be accomplished with particles that have $\beta$ (the ratio to radiation pressure force to gravitational force) $\lesssim$0.02 which is possible for a very low luminosity host star with very small ($\lesssim$0.1 $\mu$m) or very large ($\gtrsim$1 $\mu$m) dust particles. \cite{Alonso2016} observe several transits of disintegrating planetesimals around the white dwarf WD 1145+017 with the Gran Telescopio Canarias (GTC) and found no wavelength dependence in transit depth in bands centred on 0.53, 0.62, 0.71 and 0.84 $\mu$m.

We have developed a new 3D model to investigate how the optical thickness and transit cross-section of a general dust tail can affect the wavelength dependence and depth of transit light curves.   Our model builds up a tail by ejecting particles from the surface of the planet with a velocity relative to the planet and tracks them until they vanish due to sublimation, in contrast to the models of \cite{Brogi2012b}, \cite{Budaj2013}, \cite{vanWerkhoven2014}, \cite{Rappaport2014} and \cite{Sanchis-Ojeda2015} who assume a density profile in the tail.   

\cite{vanLieshout2016} release particles from the centre of the planet with zero velocity relative to the planet and without considering the effect of the planet's gravity.  To generate synthetic light curves, they calculated the individual contribution of each particle, taking into account its scattering cross-section and phase function, its extinction cross-section, and the local intensity of the stellar disk.  They then scaled these contributions by the mass-loss rate of the planet.  This limited them to only generating light curves for optically thin tails.  However, they also showed that in reality, the tail would likely have an optically thick component near the planet.

This paper is structured as follows: in Section \ref{sec:model} we describe our new model and Section \ref{sec:SimulationResults} shows the results of some instructive tail simulations.  Section \ref{sec:WavelengthDependence} explores how the wavelength dependence in transit depth depends on the optical depth of the tail, while in Section \ref{sec:tailheight} we discuss how a lower limit on particle ejection velocity can be determined from the transit depth.  Finally, Section \ref{sec:discussion} discusses the limitations and implications of the presented results and Section \ref{sec:summary} summarises the main results. 

\section{Method: The model \label{sec:model}}
\subsection{Dust dynamics code \label{subsec:pythoncode}}

Our model builds up a 3D tail by ejecting tens of thousands of meta-particles from the surface of the rocky planet, where each meta-particle represents a large number of particles.  The meta-particles can be launched in variable directions with variable speeds, allowing different launch mechanisms to be modelled.  Each individual dust meta-particle experiences a radiation pressure force away from the star, a gravitational force towards the star, and the gravitational force towards the planet.  The inclusion of the gravitational attraction of the planet means that meta-particles with low launch velocities will follow ballistic trajectories and return to the surface of the planet.      

The ratio of radiation pressure force and gravitational force towards the star, $\beta$, is independent of distance from the star and only depends on the particle's scattering properties which are determined by its composition, radius and shape \citep[e.g.][]{Burns1979}.  Our values of $\beta$ were computed as in \cite{vanLieshout2014} by integrating the radiation pressure over the spectrum of the star for a particle composition of corundum which was found by \cite{vanLieshout2016} to be consistent with the observations, however, other compositions such as iron-rich silicates are also possible.  Our simulated dust meta-particles become smaller with time due to sublimation and $\beta$ changes correspondingly. 

In reality, additional forces act on these particles, however they were neglected in this work because they produce much smaller effects.  Poynting-Robertson drag only becomes significant over many orbits \citep{vanLieshout2016} and can therefore be neglected because the lack of any correlation between transit depths implies that on average particles do not live longer than one orbit \citep{vanWerkhoven2014}. 
\cite{Rappaport2014} show in their Appendix A that the stellar wind ram pressure is expected to be one to two orders of magnitude less significant than the radiation pressure so it was also ignored in this work.  

This model was implemented in a rotating coordinate system that was centred on the star and rotated with the planet's orbital velocity so that the planet remained at the same coordinates throughout its orbit.  In this reference frame the meta-particles do not move very far over the grid in a single time step.  The equation of motion in this co-rotating reference frame is 

\begin{equation}\label{eqn:EOM}
\frac{d^2 \vec{r}}{dt^2} = -\underbrace{\frac{GM_{\star}(1-\beta(a))}{|\vec{r}|^3}\vec{r}}_{\substack{\text{stellar gravity and} \\ \text{radiation pressure}}} -\underbrace{\vphantom{\frac{\vec{x}}{|\vec{r}|^3}} 2\boldsymbol{\omega}\times\frac{d\vec{r}}{dt}}_{\text{Coriolis}} -\underbrace{\vphantom{\frac{\vec{x}}{|\vec{r}|^3}} \boldsymbol{\omega}\times (\boldsymbol{\omega}\times \vec{r})}_{\text{centrifugal}} -\underbrace{\frac{Gm_p}{|\vec{d}|^3}\vec{d}}_{\substack{\text{planetary} \\ \text{gravity}}}  
,\end{equation}

where $\boldsymbol{\omega}$ is the angular velocity vector of the planet, $\vec{r}$ is the vector from the star to the dust particle, $\vec{d}$ is the vector from the planet to the dust particle, $M_*$ is the mass of the star, and $m_p$ is the mass of the planet.  This equation of motion was integrated using Python's odeint\footnote{\url{https://docs.scipy.org/doc/scipy-0.18.1/reference/generated/scipy.integrate.odeint.html\#scipy.integrate.odeint}}, which uses Isoda from the FORTRAN library odepack. This equation of motion changes from having relatively stable solutions (non-stiff) to having potentially unstable solutions (stiff) throughout the motion of a dust particle.  The odeint package automatically determines whether an equation is non-stiff, allowing it to be accurately integrated with the fast Adams' method or stiff, requiring it to be integrated with the slower but more accurate backward-differentiation formula (BDF).

This model allows for meta-particles to be ejected with arbitrary spatial and temporal distributions so that a variety of possible ejection scenarios can be investigated, such as a spherically symmetric continuous outflow, or directed outbursts from a volcano.  However, in this work we have focussed on a simple, spherically symmetric outflow, where the meta-particle ejection direction is uniformly randomly distributed over a sphere because as was pointed out by \cite{Rappaport2012}, if the planet is tidally locked the particles might be expected to stream off the hot day-side, but if there are horizontal winds on the planet, the material could be redistributed around the planet.  

There are several important free parameters that have an impact on the tail morphology and resulting light curves.  All of these parameters are shown in Table \ref{table:ModelParameters}, along with their
typical values.

\begin{table}[]
\centering
\caption{Fiducial model input parameters.} 
\label{table:ModelParameters}
\begin{tabular}{c c}
\hline \hline 
Parameter     &  Value   \\ \hline

composition & corundum (Al$_2$O$_3$) $^{(1)}$ \\
grain density & 4.02 g cm$^{-3}$ \\
Initial meta-particle radius & 1 $\mu$m \\
Sublimation radius & 1nm \\
Particle launch direction & spherically symmetric $^{(2)}$\\
Total number of meta-particles & 5$\times10^4$ $^{(3)}$\\
Number of orbits & 1 \\
Time steps per orbit & 500 \\
Sublimation rate & -1.77$\times10^{-11}$ ms $^{-1}$ \\
Planet density & 5427 kg m$^{-3}$ $^{(4)}$ \\
Semi-major axis & 0.0131 au $^{(5)}$ \\
\hline 
Planet radius \#1& 0.0204 R$_\oplus$ $^{(6)}$  \\
Planet mass \#1  & $8.36\times10^{-6}$ M$_\oplus$\\
Planet radius  \#2 & 0.277 R$_\oplus$ \\
Planet mass  \#2 &  0.020 M$_\oplus$  \\
\hline
\multicolumn{2}{l}{Grid parameters}\\
\multicolumn{2}{l}{Radial grid}\\
Inner radius & 0.0130 au \\
Outer radius & 0.0150 au \\
Bin size & 1.50$\times10^6$ \\
\multicolumn{2}{l}{Elevation grid ($0^\circ - 180^\circ$)} \\
Lower elevation & 89$^\circ$ $^{(7)}$  \\
Upper elevation & 91$^\circ$ $^{(8)}$ \\
Bin size & 0.0526$^\circ$ \\
\multicolumn{2}{l}{Azimuthal grid ($0^\circ - 360^\circ$)} \\
Bin size & 0.5$^\circ$ \\
\hline
\multicolumn{2}{l}{$^{(1)}$ From \cite{vanLieshout2016}.}\\
\multicolumn{2}{l}{$^{(2)}$ Since the dust may be subject to horizontal winds on the}\\
\multicolumn{2}{l}{planet that can distribute material from the substellar point}\\
\multicolumn{2}{l}{to the night side \citep{Rappaport2012}.}\\
\multicolumn{2}{l}{$^{(3)}$ The number of particles that each meta-particle represented} \\
\multicolumn{2}{l}{was scaled to set the planet's mass-loss rate to the desired value.} \\
\multicolumn{2}{l}{$^{(4)}$ Equal to the bulk density of Mercury. }\\
\multicolumn{2}{l}{$^{(5)}$ See footnote \ref{Footnote:SemiMajorAxis}.}  \\
\multicolumn{2}{l}{$^{(6)}$ The upper-limit determined} \\
\multicolumn{2}{l}{by \cite{vanWerkhoven2014} is 0.7 R$_\oplus$.} \\
\multicolumn{2}{l}{$^{(7)}$ With an additional large bin containing $0^\circ - 89^\circ$.}\\
\multicolumn{2}{l}{$^{(8)}$ With an additional large bin containing $91^\circ - 180^\circ$.} \\

\end{tabular}
\end{table}

After the meta-particles are ejected from the planet, they sublimate until they reach a radius of 1 nm and are removed from the simulation.  We assumed a simple sublimation rate that was constant for all meta-particles and over all meta-particle radii.  In reality, the sublimation rate would be more complex and would depend on the compositions, shapes and temperatures of the particles which was partially exploited by \cite{vanLieshout2014, vanLieshout2016} to constrain the particle composition.  However, for this work our focus was on investigating how the transit depth varied as a function of wavelength and meta-particle ejection velocity for a general tail, so our only requirements on the sublimation rate were that it produced a tail of reasonable length and that meta-particles did not survive for longer than one orbit (since there is no correlation between consecutive transit depths \citep{vanWerkhoven2014}), making our simple approximation reasonable.

Our model continuously ejects a stream of meta-particles so that at every time step of the simulation, there are several thousand spatially separated meta-particles populating the tail.  This enables us to investigate whether the optical depth in the radial direction through the tail can reduce the flux (and radiation pressure) on shielded dust meta-particles enough to affect the tail's morphology.  However, that is beyond the scope of this paper and will be presented in a forthcoming paper.

Our planet properties were chosen in the following way.  The trialled planet mass of $8.36\times10^{-6}$ M$_\oplus$ (\#1) was chosen by trial-and-error so that the planet's gravity would have a very small effect on the meta-particles' motion and the trialled planet mass of 0.02 M$_\oplus$ (\#2) was chosen because it was found by \cite{Perez-Becker2013} to be its most likely current mass. The planet's bulk density was chosen to be equal to that of Mercury because that assumption has been previously made \citep[e.g.][]{Perez-Becker2013}.  This density was used to calculate the radii corresponding to the trialled masses, assuming a spherical planet.

In all of our tail simulations, we used a constant initial meta-particle size instead of a distribution.  This was primarily for simplicity because the radiative transfer component of our model (see Section \ref{sec:raytracing}) is too slow to allow the model parameters to be constrained in a Markov-chain Monte-Carlo (MCMC) manner.  We chose an initial meta-particle size of 1 $\mu$m as this was generally consistent with the findings of previous studies (see Introduction) and also with dynamical constraints discussed in Section \ref{sec:lareparticles}.

\subsection{Particle dynamics simulations \label{sec:ModelValidation}}

\begin{figure*}[h] 
\centering 
\includegraphics[width=20cm]{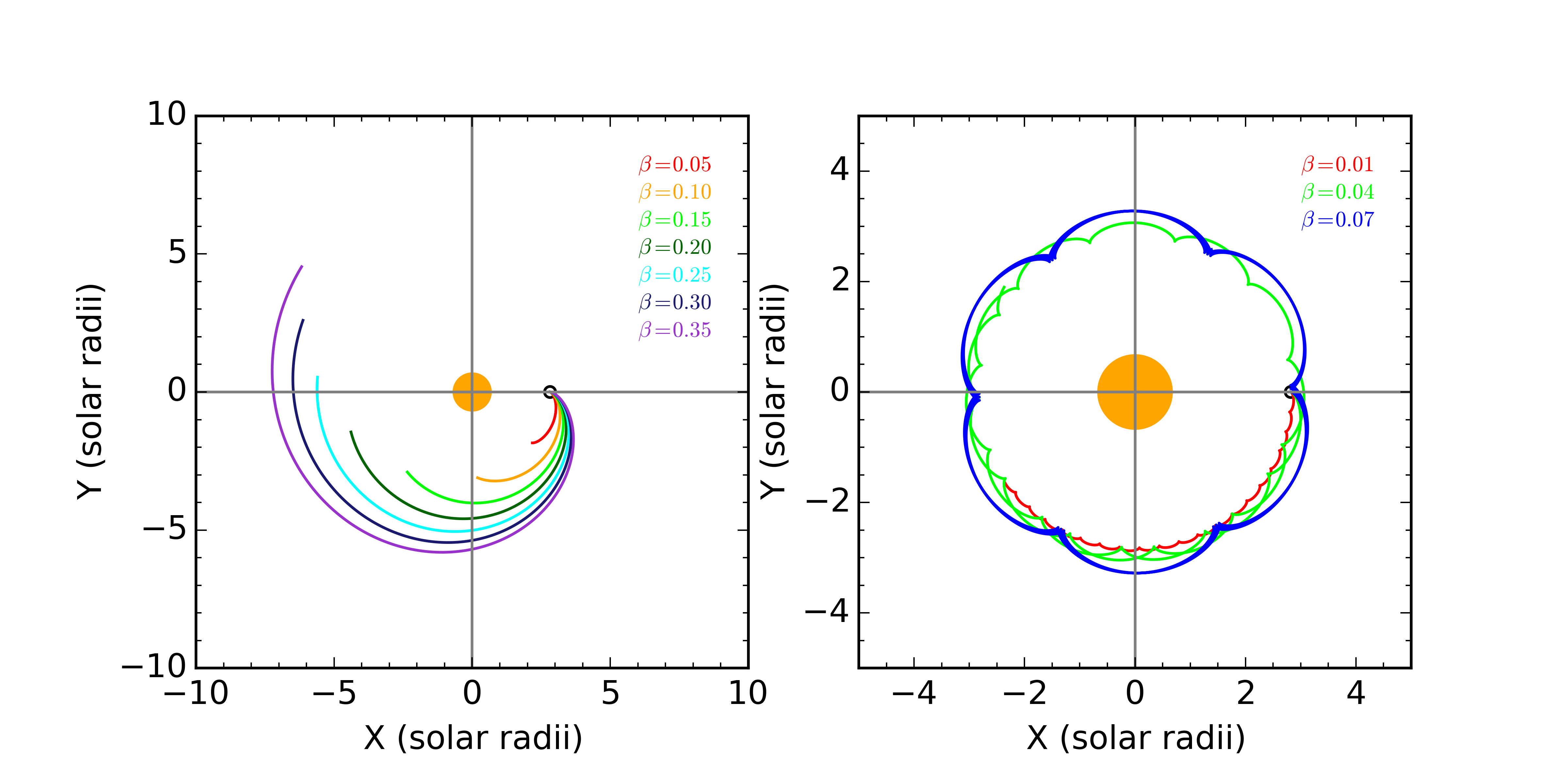} 
\caption{Reproduction of the particle tracks shown in Fig. 7 of \cite{Rappaport2014} to validate the accuracy of our particle dynamics code.  Both panels show the tracks of non-sublimating particles in the corotating frame of the planet.  Left: Tracks of particles after one planetary orbit for radiation pressure force to gravitational force ratios, $\beta$, that vary from 0.05 to 0.35.  Right: Same as left but for 20 planetary orbits, with $\beta =$ 0.01, 0.04 and 0.07.  The cusps are the periastron passages of the dust particles.  The orange circle represents the approximate size of the host star, Kepler-1520.}
\label{fig:rosettas} 
\end{figure*}

To validate our code, we studied the tracks of non-sublimating particles, which have a constant $\beta$, that were released from the planet centre with zero velocity relative to the planet.  The values of $\beta$ were such that the particles stayed in bound orbits, which is true for $\beta < 0.5$ \citep{Rappaport2014}.  Such bound particles should form rosette-like shapes in the co-rotating frame over many orbits, as is shown in Fig. 2 of \cite{vanLieshout2016} and Fig. 7 of \cite{Rappaport2014}.  We reproduce Fig. 7 of \cite{Rappaport2014} in our Fig. \ref{fig:rosettas}, showing perfect agreement and hence confirming that the numerical accuracy of our dynamics code was sufficient to reliably solve the equation of motion describing the motion of the particles. 

\begin{figure*}[h] 
\centering 
\includegraphics[width=20cm]{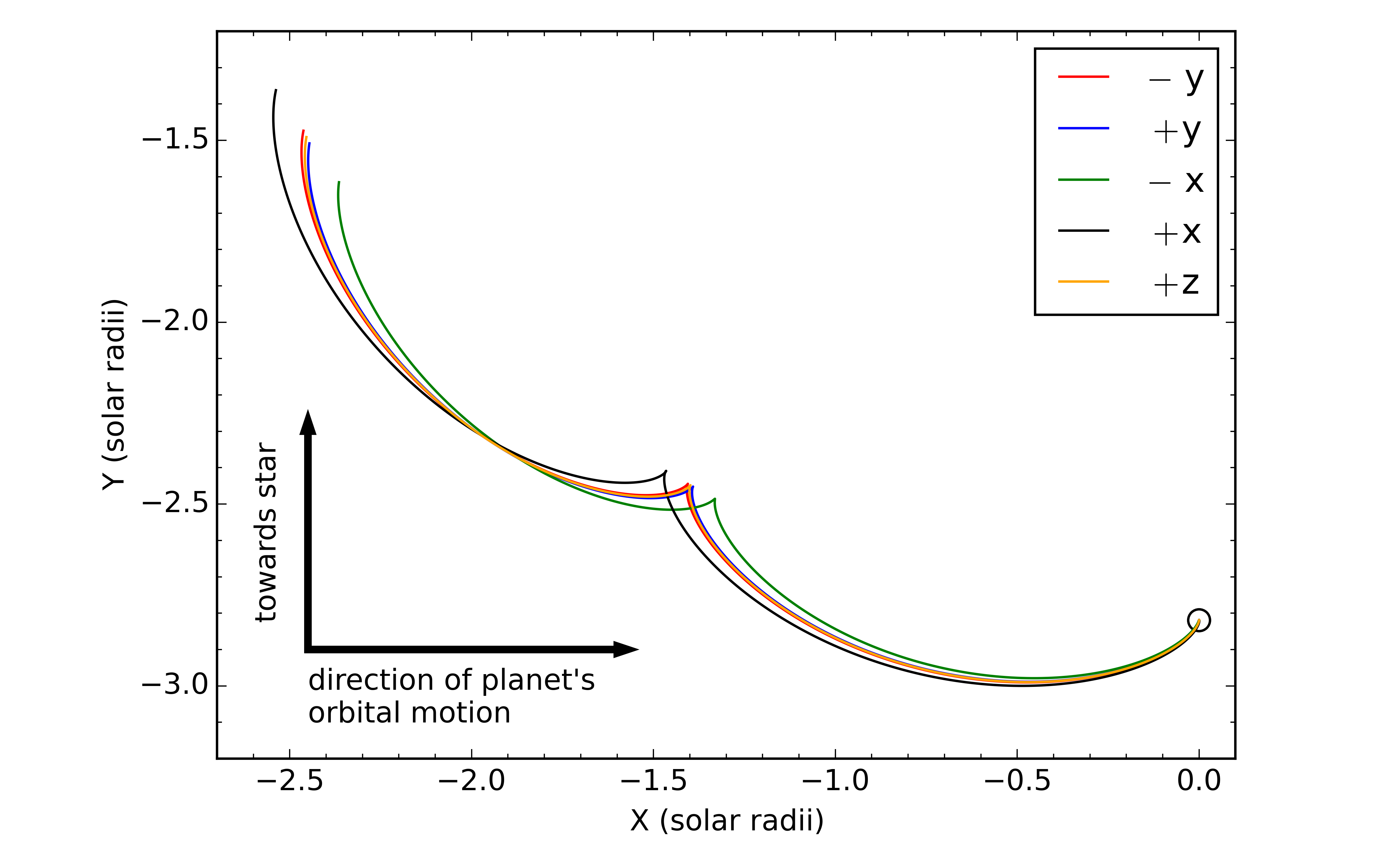} 
\caption{Trajectories after one planetary orbit of non-sublimating particles of corundum of radius 1 $\mathrm{\mu}$m with $\beta = 0.038$ after being ejected in different directions from the surface of a planet of radius 0.020 R$_\oplus$ and mass 8.4$\times10^{-6}$ M$_\oplus$.  The particles were ejected with a velocity of 1.2 times the surface escape velocity (272 ms$^{-1}$) towards the star at the top of the page (+y) , in the anti-stellar direction towards the bottom of the page ($-$y), in the direction of the planet's orbital motion to the right of the page (+x), in the anti-orbital motion direction to the left of the page ($-$x) and perpendicular to the orbital plane, out of the page (+z).  This coordinate system is rotating around the z axis so the track for $-$z is the same as for $+$z.  The planet is indicated by the circle (not to scale).}
\label{fig:RosettasWithV} 
\end{figure*} 

If the particles are ejected from the surface of the planet with some velocity relative to the planet, the track of each particle differs from the track produced by releasing the particles from the centre of the planet.  This is shown in Fig. \ref{fig:RosettasWithV} which shows the tracks of spherical particles of corundum of radius 1 $\mu$m, with $\beta = 0.038$.  It can be seen in Fig. \ref{fig:rosettas} that when all the particles are released from the centre of the planet with no relative velocity, the perihelion point forms a cusp for all particles.  However, when the particles are ejected from the surface of a planet of mass 8.4$\times10^{-6}$ M$_\oplus$ and radius 0.020 R$_\oplus$ with a velocity of 1.2 times the surface escape velocity (272 ms$^{-1}$) the perihelion point is not the same for all particles and depends on the ejection velocity.  This causes the local enhancement in density at the perihelion cusp to be spread slightly along the planet's orbit. 

To ensure that our constant time steps were small enough to enable Eq. \ref{eqn:EOM} to be accurately solved, we doubled the number of time steps, which changed the average displacement between individual meta-particles by less than 0.5 planetary radii (assuming a planet radius of 0.28 R$_\oplus$).  This is negligible compared to the size of the tail, which has a maximum extent perpendicular to the planet's orbital plane of 10 $-$ 20 planetary radii and typical length of 1000 planetary radii.

\subsection{Ray tracing with MCMax3D \label{sec:raytracing}}

The code described in Section \ref{subsec:pythoncode} simulates the dynamics of the dust meta-particles in the tail but does not generate light curves.  To generate light curves, we employed the radiative transfer code MCMax3D\footnote{\url{http://www.michielmin.nl/codes/mcmax3D/}} \citep{Min2009}.  MCMax3D was originally designed to generate circumstellar disk density distributions and carry out Monte Carlo radiative transfer.  We modified MCMax3D, to take an arbitrary mass density distribution file as an input.  The code described in Section \ref{subsec:pythoncode} converts the distribution of individual meta-particles to a continuous mass density distribution for MCMax3D.  The density is calculated on a spherical grid surrounding the star that has cell dimensions that were chosen so that there were always several meta-particles per cell and that the distribution was always continuous, without unpopulated cells between populated cells.  This density grid was also used for the radiative transfer, and consisted of 200 evenly spaced bins in the radial direction ranging from 0.0130 $-$  0.0150 au from the centre of the star (with the fiducial semi-major axis of Kepler-1520 b being 0.0131 au\footnote{\label{Footnote:SemiMajorAxis} This value differs from the value given by \cite{vanWerkhoven2014} of 0.0129(4) au because our value was derived by solving Kepler's third law with an orbital period of 15.685 hours \citep{Rappaport2012} and assuming a stellar mass of 0.704 $M_\odot$ which is only approximately the value found by \cite{Huber2014} of 0.666 $M_\odot$.}), 720 evenly spaced bins in the azimuthal direction, ranging from 0 to 360$^{\circ}$, and 40 bins of elevation angle ranging from 0 to 180$^{\circ}$ (where the planet's orbital plane is at 90$^{\circ}$), with the first bin containing 0$-$89$^{\circ}$, the last bin containing 91$-$180$^{\circ}$ and the remaining 2$^{\circ}$ close to the orbital plane being covered by 38 evenly spaced bins.  The grid boundaries were set such that the planet fell on an intersection of grid lines so that meta-particles released from different sides of the planet would be in different grid cells.  Since the 3D spherical grid completely surrounded the star, most of the grid cells were empty, however some cells contained mass, distributed in the same way as the tail produced by the code in Section \ref{subsec:pythoncode}.

The MCMax3D code was then used to carry out a full 3D radiative transfer through this grid by propagating $1\times10^6$ photons though the mass density distribution in a Monte Carlo fashion with photons being emitted from the star at all angles.  We used a full treatment of scattering that includes extinction due to scattering by using the distribution of hollow spheres (DHS) method from \cite{Min2005}, which is analogous to Mie scattering but is more general as it can be applied to non-spherical particles.  To produce images, the simulated photons were detected by a virtual camera situated such that photons would propagate from the star, through the dust, before being detected and producing an image composed of photons from all angles from the stellar disk.

We assumed that the dust particles in the tail were composed of corundum (Al$_2$O$_3$) as this was determined by \cite{vanLieshout2016} to be consistent with the observations of Kepler-1520 b (although other compositions are possible).   We took the optical properties of corundum from \cite{Koike1995} and constructed the opacities by assuming irregularly shaped particles, using the DHS method.  The opacity as a function of grain size, integrated over the spectrum of Kepler-1520 is shown in Fig. 2 of \cite{vanLieshout2014}. 

The virtual camera was elevated relative to the orbital plane to approximate the transit's impact parameter.  This was only an approximation because there is a slight mismatch between the effective impact parameter derived for Kepler-1520 b in previous research (see Introduction) and the viewing elevation used here because different parts of the tail are at slightly different radial distances from the host star.  Light curves were generated by rotating the virtual camera around the system to mimic the effect of having a stationary observer observing a transiting dust tail.

\begin{figure*}[h] 
\centering 
\includegraphics[width=18cm]{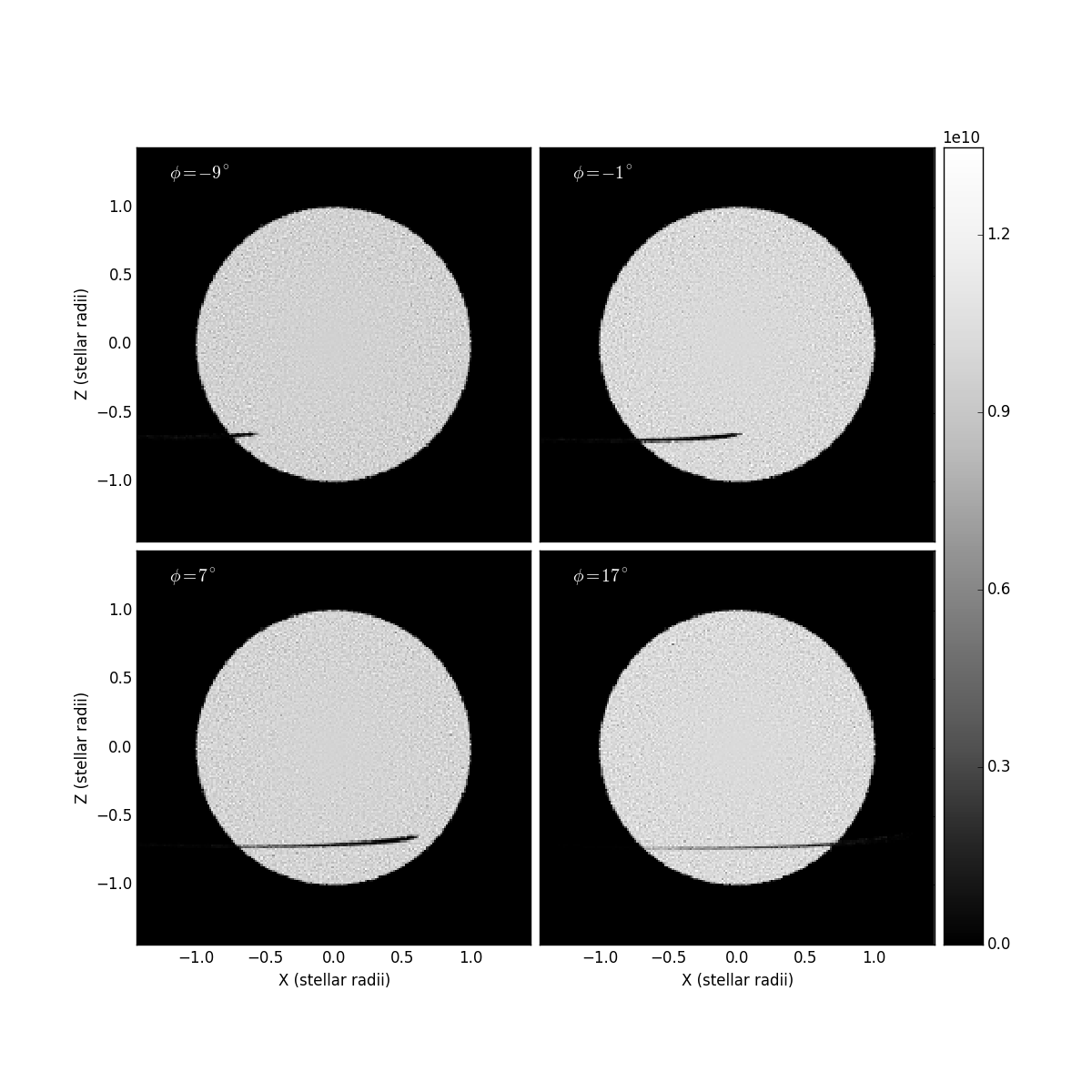} 
\caption{Images generated by MCMax3D for the tail configuration presented in Section \ref{sec:OpticallyThin} at $\lambda = 650$ nm for different azimuthal viewing angles corresponding to orbital phases $\phi = -9^\circ$, $-1^\circ$, $7^\circ$ and $17^\circ$, with an elevation viewing angle of 81.52$^\circ$ as measured from the pole of the orbital plane.  Integrating the flux of images such as these for different azimuthal viewing angles produces a transit light curve.} 
\label{fig:MCMax3DExample} 
\end{figure*}

\begin{figure*}[h] 
\centering 
\includegraphics[width=12cm]{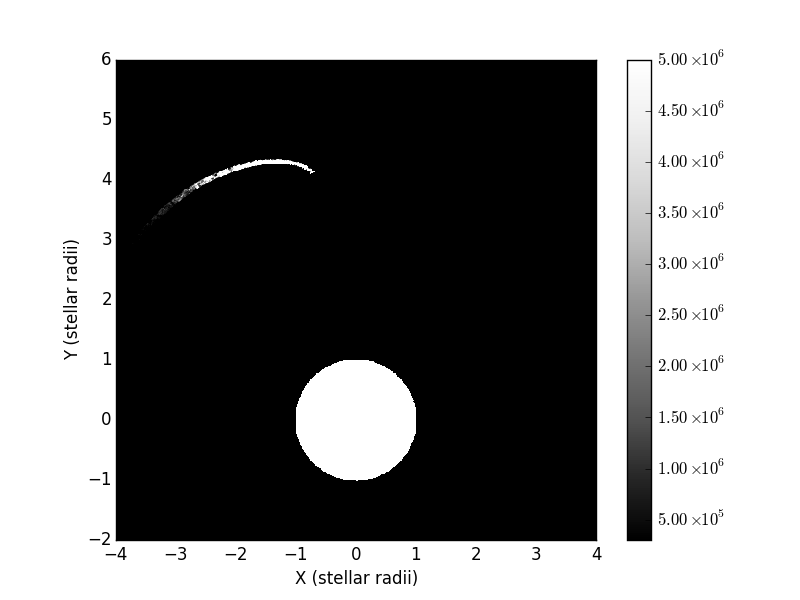} 
\caption{Same as Fig. \ref{fig:MCMax3DExample},  for an elevation viewing angle of $90^\circ$, looking down on the orbital plane of the planet.  The tail scatters little light vertically out of the orbital plane so the dynamic range has been restricted to the fluxes from the brightest and faintest parts of the tail so the star is actually three orders of magnitude brighter than indicated by the upper limit of this colour scale.} 
\label{fig:MCMax3DTopDown} 
\end{figure*}

Examples of these simulated images are shown in Figs. \ref{fig:MCMax3DExample} and \ref{fig:MCMax3DTopDown}.  Figure \ref{fig:MCMax3DExample} shows a series of images at different orbital phases from a viewing elevation of 81.52$^\circ$ from the pole of the orbital plane (approximating the impact parameter), while Fig. \ref{fig:MCMax3DTopDown} shows an image at a single orbital phase as viewed from the pole of the orbital plane, with elevation 180$^\circ$.

Images (and hence transit light curves) can be generated in different wavelengths, which allows the wavelength dependence of the transit depth to be studied.  MCMax3D is also capable of modelling polarisation, allowing us to predict the degree of polarisation, $\sqrt{Q^2+U^2}/I$, induced by the dust in the tail.  

The ray-tracing carried out by MCMax3D is computationally very intensive and takes about 15 minutes to generate a single image at a single wavelength on a standard desktop workstation.  To generate a light curve of sufficiently high temporal resolution, images for a large number of viewing angles need to be generated (e.g. 360 viewing angles for a 1$^\circ$ orbital phase resolution, corresponding to a temporal resolution of 157 s), so the time required to generate a full phase light curve for a single wavelength is typically 80 hours. For this reason, when simulating full light curves, we consider only the wavelengths 0.65 $\mu$m (Kepler bandpass), 0.85 $\mu$m and 2.5 $\mu$m.  When only the transit depth from a single viewing angle was needed, we considered the wavelengths of 0.45 $\mu$m, 0.65 $\mu$m, 0.85 $\mu$m, 1.5 $\mu$m, and 2.5 $\mu$m.

%\clearpage

\section{Results of simulations \label{sec:SimulationResults}}

\subsection{Modelling the light curve of Kepler-1520 b with a low planet mass \label{sec:OpticallyThin}}

By keeping most of the parameters fixed (see Section \ref{subsec:pythoncode}), we were able to vary the meta-particle ejection velocity and dust tail mass in a trial-and-error way to produce a reasonable match to the observed average Kepler  long cadence (LC) light curve of Kepler-1520 b (although this may not be the best match that this model can produce).  The meta-particle ejection velocity set the tail's maximum extent perpendicular to the planet's orbital plane (which is proportional to its transit cross-section) and its mass determined its opacity.

To produce this tail, we used a planet mass of $8.36\times10^{-6}$ M$_\oplus$ and radius 0.0204 R$_\oplus$ (mass and radius \#1 in Table \ref{table:ModelParameters}), which is much smaller than the limit of 0.7 R$_\oplus$ determined by \cite{vanWerkhoven2014} and would give a transit depth of 8$\times$10$^{-6}$\%.  Meta-particles were ejected with a velocity of 680 ms$^{-1}$ (three times faster than the surface escape velocity).  This resulted in a maximum tail height above the orbital plane of $1.3\times10^7$ m.  This tail was mostly optically thin, however, it was moderately optically thick at the head of the tail, close to the planet. For this transit cross-section, we found that a dust tail mass of $4.8\times10^{13}$ kg was required to produce a relatively good match to the Kepler average long-cadence light curve.  This corresponds  to a mass-loss rate of 18.8 M$_\oplus$Gyr$^{-1}$. 

\begin{figure*}[h] 
\centering 
\includegraphics[width=18cm]{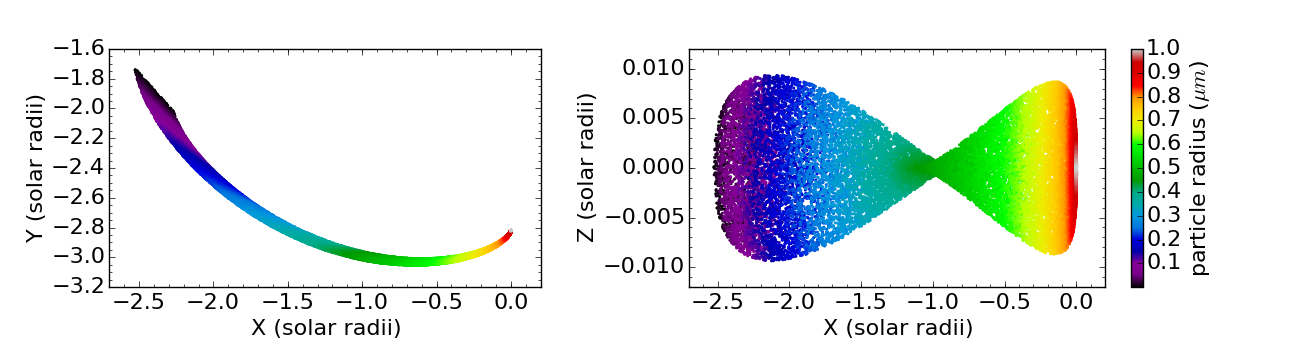} 
\caption{Simulated tail of corundum meta-particles viewed from above the orbital plane (left) and in the orbital plane (right) with meta-particles colour coded according to meta-particle radius.  The left panel's axes have the same scale, however the right panel's vertical axis is stretched  by a factor of $\sim$1000 relative to the horizontal axis because the tail is much longer than it is high.  These meta-particles have an initial radius of 1 $\mu m$ and were ejected with a spherically symmetric distribution from a planet of mass 8.4$\times10^{-6}$ M$_\oplus$ and radius of 0.020 R$_\oplus$ at a velocity of 3.0 times the planet's surface escape velocity (or 674 ms$^{-1}$).  The meta-particles were tracked as they sublimated until they were removed when they reached a radius of 1 nm.  The sublimation rate of the meta-particles was set so that they reached a radius of 1 nm after one planetary orbit.} 
\label{fig:1780_radius} 
\end{figure*}

\begin{figure*}[h] 
\centering 
\includegraphics[width=18cm]{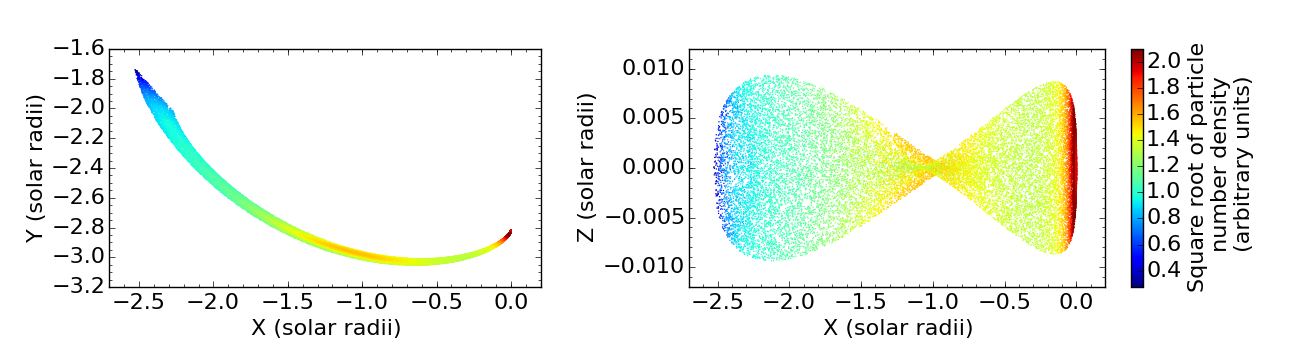} 
\caption{Same as Fig. \ref{fig:1780_radius} but colour coded proportionally to the square root of density to increase the dynamic range.} 
\label{fig:1780ScatterDens} 
\end{figure*}

Visualisations of this tail are shown in Figs. \ref{fig:1780_radius} and \ref{fig:1780ScatterDens} which show the tail with meta-particles colour coded according to meta-particle radius and the square root of the density, respectively.
This tail has a smooth morphology and the perihelion point where all of the meta-particles on inclined orbits cross back through the orbital plane of the planet can be clearly seen as a waist in the `bow-tie' plot of Figs. \ref{fig:1780_radius} and \ref{fig:1780ScatterDens}.  The points in Fig. \ref{fig:1780ScatterDens} are colour coded according to the square root of the density (to increase the dynamic range) and clearly show a local density enhancement at this perihelion point.  This enhancement has interesting implications for tails with a high optical depth, as shown in Section \ref{sec:doubledip}. 

Even though we ejected dust meta-particles with a constant radius of 1 $\mathrm{\mu}$m, a distribution of meta-particle sizes in the tail is produced by the meta-particles sublimating.  We used a constant sublimation rate which leads to the distribution in the tail as a whole being described by a power-law of the form $dN/da \propto a^{-\Gamma}$ where $a$ is the meta-particle radius and $\Gamma = 1$. This value of $\Gamma$ is different to the value used in \cite{Brogi2012b} of $\Gamma = 3.5$, however, it is broadly consistent with the range of values derived by \cite{Sanchis-Ojeda2015} for the dust tail of K2-22 b of $\Gamma = 1 - 3$.

\begin{figure*}[h] 
\centering 
\includegraphics[width=16cm]{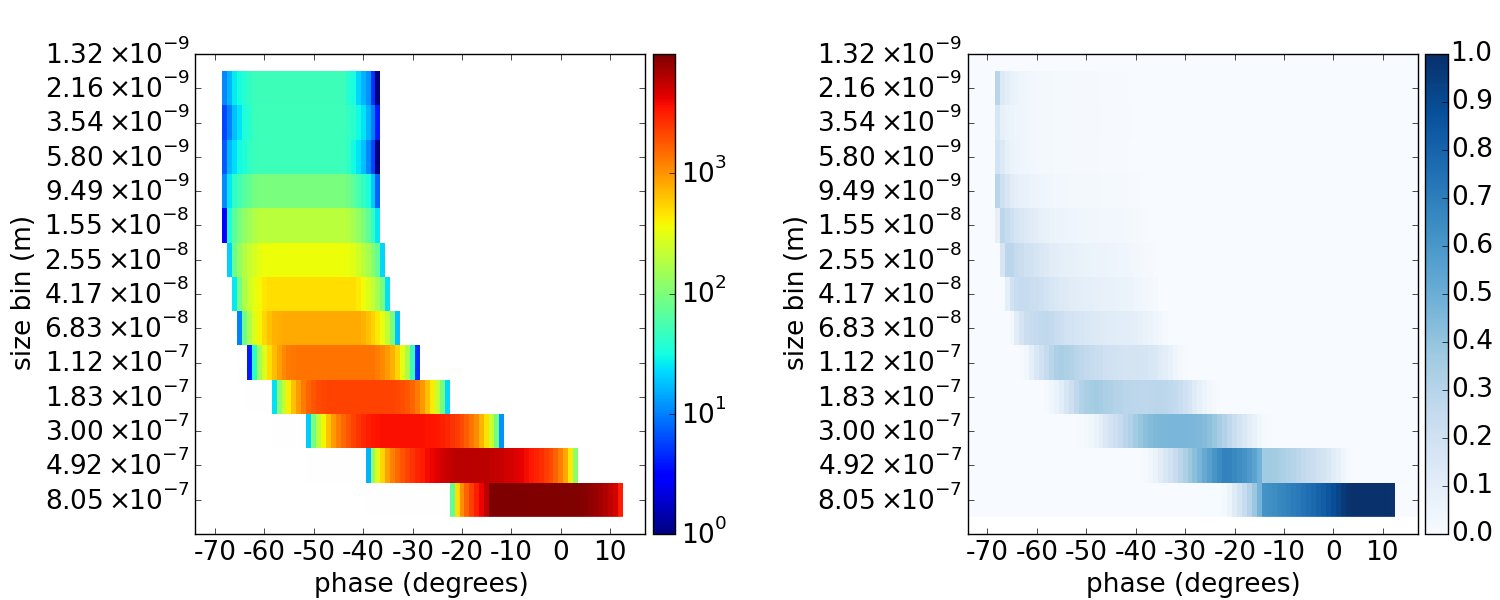} 
\caption{Distribution of meta-particle sizes in the tail produced by ejecting meta-particles with a spherically symmetric distribution from a planet of mass  8.4$\times10^{-6}$ M$_\oplus$ and radius of 0.020 R$_\oplus$ at a velocity of 3.0 times the planet's surface escape velocity (or 674 ms$^{-1}$).  Left: Number of meta-particles in each size bin (vertical axis) as a function of angular displacement along the tail with positive phases being ahead of the planet (horizontal axis).  Right: Same as left but instead of showing the absolute number of meta-particles, it shows the probability of finding a meta-particle within a given size bin at that angular displacement along the tail, so that the sum over all meta-particle sizes for a given angular phase (column) is one.} 
\label{fig:LC1780ParticleDist} 
\end{figure*}

To consider the meta-particle size distribution in more detail, the distribution of meta-particle sizes as a function of phase along the tail is shown in Fig. \ref{fig:LC1780ParticleDist}.  The left panel shows the number of meta-particles in each size bin and the right panel shows the probability of finding a meta-particle of a given size.  This shows the details of how the number and size of meta-particles decreases with increasing angular distance away from the planet.  

\begin{figure*}[h] 
\centering 
\includegraphics[width=14cm]{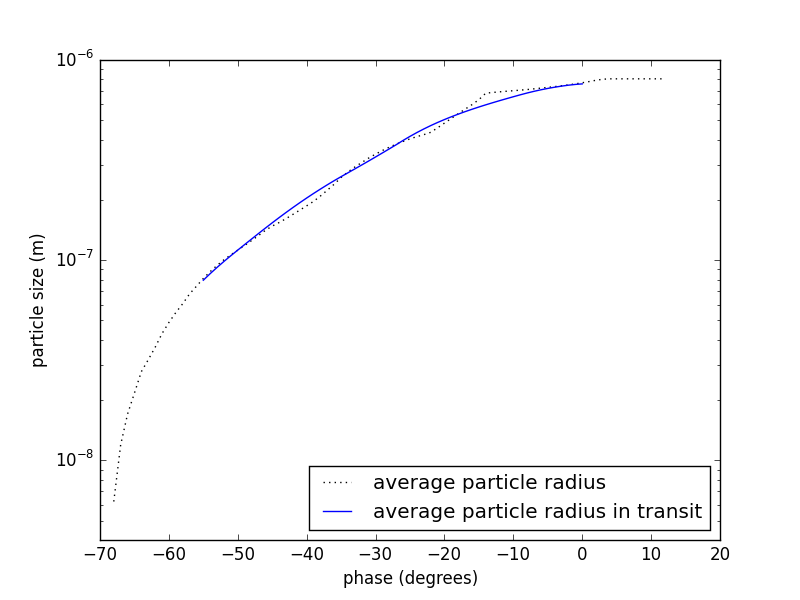} 
\caption{Average meta-particle size in transit for the tail produced by ejecting meta-particles with a spherically symmetric distribution from a planet of mass 8.4$\times10^{-6}$ M$_\oplus$ and radius of 0.020 R$_\oplus$ at a velocity of 3.0 times the planet's surface escape velocity (or 674 ms$^{-1}$).  The black dotted line shows the average meta-particle size as a function of phase in transit in intervals of $1^\circ$ while the solid blue line is convolved by the angular size of Kepler-1520 as seen from Kepler-1520 b, of 26$^\circ$, to show that at different times during the transit, different meta-particle sizes dominate the contribution to the light curve.  This excludes the meta-particles external to the stellar disk that contribute with scattered starlight.} 
\label{fig:LC1780AvgInTransit} 
\end{figure*}

Therefore, the distribution of meta-particles contributing to the transit light curve changes as a function of orbital phase.  Fig. \ref{fig:LC1780AvgInTransit} shows the average size of meta-particles crossing the stellar disk  as a function of orbital phase during transit.  These meta-particles make the most significant contribution to the transit light curve, however they are not the only contribution to the light curve because meta-particles not in front of the stellar disk can also contribute to the light curve by forward-scattering light.  However, it is clear that even if the meta-particles are ejected with a constant initial meta-particle size, the combination of the meta-particles sublimating and having different ejection velocities will result in a distribution of meta-particle sizes as a function of orbital phase.

\begin{figure*}[h] 
\centering 
\includegraphics[width=16cm]{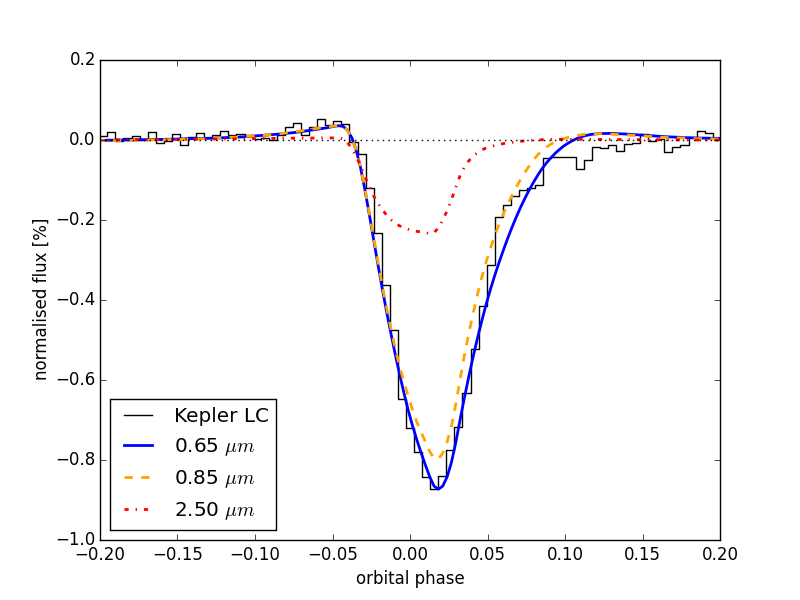} 
\caption{Model light curves produced by the tail shown in Fig. \ref{fig:1780_radius} at wavelengths of 0.65 $\mu$m (solid blue), 0.85 $\mu$m (dashed orange) and 2.5 $\mu$m (dot-dashed red) compared with the Kepler long-cadence light curve of Kepler-1520 b (black).  The model light curves are convolved to the Kepler long-cadence of 30 minutes.  To produce this light curve, meta-particles were ejected with a spherically symmetric distribution from a planet of radius 0.020 R$_\oplus$ and mass 8.4$\times10^{-6}$ M$_\oplus$ with a velocity of 3.0 times the planet's surface escape velocity (or 674 ms$^{-1}$) at a mass loss rate of 18.8 M$_\oplus$Gyr$^{-1}$.}
\label{fig:LC1780} 
\end{figure*}

The transit light curve for wavelengths of 0.65, 0.85 and 2.5 $\mu$m that this tail produced are shown in Figure \ref{fig:LC1780}.  We compare these simulated light curves to the Kepler LC light curve of Kepler-1520 b that resulted from the de-correlation and de-trending of 15 quarters of Kepler data by \cite{vanWerkhoven2014}.  

The 0.65 $\mu$m light curve (Kepler bandpass) is very similar to the Kepler LC light curve, with the pre-ingress forward-scattering peak, ingress and egress slopes and transit width matching the Kepler LC data reasonably well.  It can be seen that at this dust mass-loss rate, the transit light curve depends significantly on wavelength with a large difference in transit depth and shape from the visible to the near infrared.  This difference may even be able to constrain the mass loss rate, as will be discussed in Section \ref{sec:WavelengthDependence}.    

We computed the light curve at 0.65 $\mu$m over the entire orbital phase to search for signs of a secondary eclipse but no secondary eclipse was apparent.  This is consistent with the Kepler LC observations  \citep{vanWerkhoven2014}.

\subsection{Optically thick tail \label{sec:doubledip}}

\begin{figure*}[h] 
\centering 
\includegraphics[width=14cm]{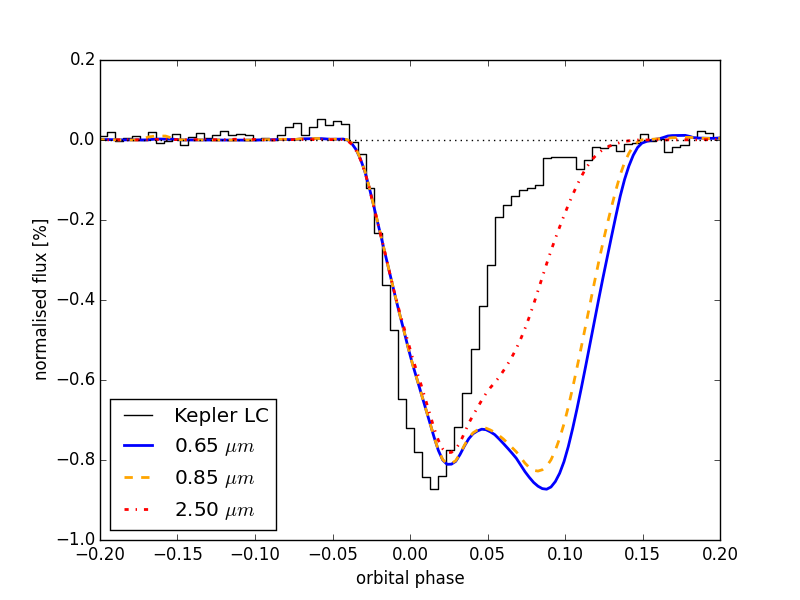} 
\caption{Model light curves produced by the optically thick tail described in Section \ref{sec:doubledip} at wavelengths of 0.65 $\mu$m (solid blue), 0.85 $\mu$m (dashed orange) and 2.5 $\mu$m (dot-dashed red) compared with the Kepler long-cadence light curve of Kepler-1520 b (black).  The model light curves are convolved to the Kepler long-cadence of 30 minutes.  This tail was produced by ejecting particles with a spherically symmetric distribution from a planet of mass 8.4$\times10^{-6}$ M$_\oplus$ and radius of 0.020 R$_\oplus$ at a velocity of 413 ms$^{-1}$) (1.8 times the planetary surface escape velocity) and scaling its final dust mass to $1.2\times10^{16}$ kg, or a dust mass-loss rate of 4.8$\times10^3$ M$_\oplus$ Gyr$^{-1}$ (600 times higher than the tail mass that produced Fig. \ref{fig:LC1780}) to make it completely optically thick.}
\label{fig:LC1826} 
\end{figure*}

To investigate the properties of a hypothetical optically thick tail, we produced a tail of dust that had 600 times more mass than the tail presented in Section \ref{sec:OpticallyThin}, giving a dust mass of $1.2\times10^{16}$ kg, or a dust mass-loss rate of 4.8$\times10^3$ M$_\oplus$ Gyr$^{-1}$.  This planet dust mass-loss rate is unrealistically high because the planet would not survive for long enough to have a reasonable chance of being observed.  However, it produces interesting light curves, so we present it here as a hypothetical illustrative example. We see four major differences when comparing its light curves shown in Fig. \ref{fig:LC1826} to those shown in Fig. \ref{fig:LC1780}.  The first difference is that the transit duration has become longer because the small meta-particles in the low density region at the end of the tail now have enough mass to make an appreciable effect on the light curve.  The second difference is that the wavelength dependence in transit depth has become much less significant, while the third difference is that the pre-ingress forward-scattering feature is no longer present.  The fourth difference is the `double dip' transit shape which results from the tail being bow-tie shaped (right panel of Fig. \ref{fig:1780_radius}) and optically thick.  As was previously discussed by \cite{vanLieshout2016}, this occurs because when the particles on inclined orbits pass through the planet's orbital plane (the narrow part of the bow-tie), they leave gaps above and below the orbital plane, reducing the tail's cross-section.  The absorption of an optically thick tail only depends on the tail cross-section so this reduces the absorption at the mid-transit point.  Therefore, if such a feature were ever observed in the light curve of a disintegrating planet, it would indicate that the tail was optically thick and that the particles were necessarily surviving for at least half of an orbit to reach this point of tail cross-section reduction.  

By exploiting the fact that the mid-transit depth depends linearly on maximum tail height for an optically thick tail, this transit depth was made to be comparable to the depth of the average Kepler LC light curve of Kepler-1520 b by setting the meta-particle ejection velocity to be 413 ms$^{-1}$ (1.82 times the surface escape velocity) which resulted in a maximum tail height from the orbital plane of $7.38\times10^{6}$ m.  The relation between meta-particle ejection velocity and maximum tail height is further discussed in Section \ref{sec:tailheight}.

We checked for a secondary eclipse at 0.65 $\mu$m with this tail mass by computing the full orbit phase curve.  However, as with the optically thin tail, we did not see a secondary eclipse.  While this is an interesting illustrative example, this scenario is unlikely.

\subsection{Modelling the light curve of Kepler-1520 b with a planet mass of 0.02 M$_\oplus$ \label{sec:TailPlanetGrav1}}

\begin{figure*}[h] 
\centering 
\includegraphics[width=19cm]{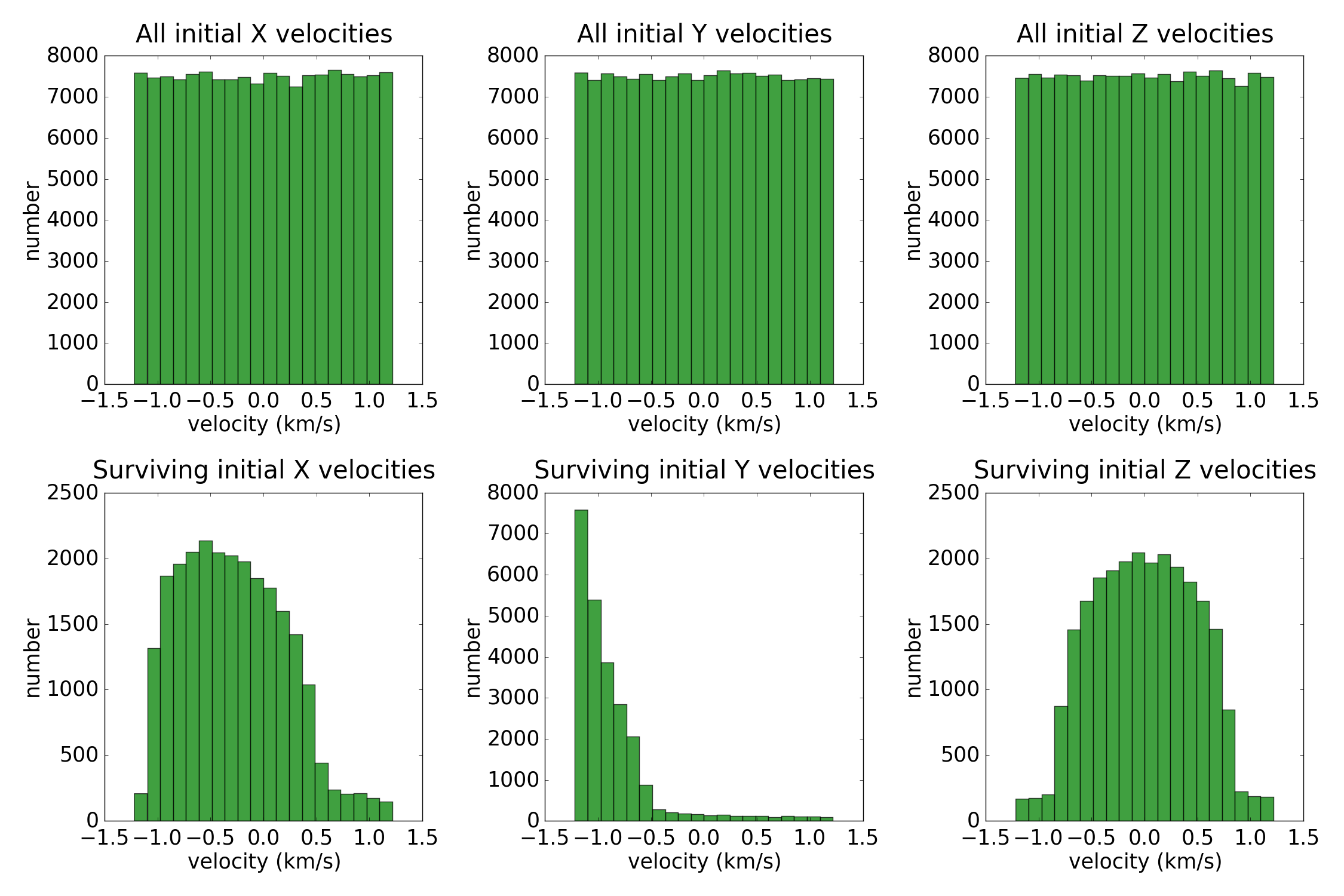} 
\caption{Initial velocity components of all meta-particles after being ejected from a planet of mass 0.02 M$_\oplus$ in a continuous and spherically symmetric distribution with a velocity of 0.40 times the surface escape velocity (or 1.21 kms$^{-1}$) (top) and only the particles that do not collide with the planet and ultimately form a tail (bottom). The components are: in the orbital direction, $\vec{X}$ (left), in the stellar direction, $\vec{Y}$ (middle), and normal to the orbital plane, $\vec{Z}$ (right).} 
\label{fig:SurvivingInitialVelocityDistribution} 
\end{figure*}

Since the planet of mass $8.36\times10^{-6}$ M$_\oplus$ that was used in Section \ref{sec:OpticallyThin} would disintegrate too quickly, we simulated a tail using a planet mass of 0.02 M$_\oplus$ (mass \#2 in Table \ref{table:ModelParameters}) and a meta-particle ejection velocity of 1.21 km s$^{-1}$ or 0.40 times the surface escape velocity.  This resulted in a maximum height from the orbital plane of 1.5$\times10^{7}$ m, which is similar to the maximum height of the tail presented in \ref{sec:OpticallyThin}, however the maximum height is just an approximate comparison between these tails because they have different vertical meta-particle distributions.  Ejecting the meta-particles at such a low velocity resulted in 84\% of the meta-particles falling back onto the planet in ballistic trajectories before they could form a tail.  To compensate for this large number of lost meta-particles, we increased the number of ejected meta-particles so that the final number of meta-particles was the same as the tail shown in Section \ref{sec:OpticallyThin}.  The surviving 16\% of meta-particles have an interesting distribution of initial velocities which is shown in Fig. \ref{fig:SurvivingInitialVelocityDistribution} where the upper panels show the distribution of initial velocities of all ejected meta-particles and the lower panels show the initial velocity distribution of only the meta-particles that do not collide with the planet and ultimately form a tail.   The directional components are: in the direction of the planet's orbital motion ($\vec{X}$), directed towards the star ($\vec{Y}$), and directed perpendicular to orbital plane ($\vec{Z}$). There is a strong preference for tail forming meta-particles to have been ejected in the anti-orbital direction, the anti-stellar direction and at small angles from the planet's orbital plane.

Meta-particles that are ejected in the anti-orbital direction are more likely to avoid colliding with the planet than meta-particles that are ejected in the orbital direction because the radiation pressure and centrifugal force act to move the meta-particles radially away from the star, slowing their orbital velocity and allowing them to be overtaken by the planet.  In the co-rotating reference frame, the meta-particles drift away from the stationary planet in the anti-orbital direction.  Therefore meta-particles ejected in the orbital direction have to pass over the planetary surface, increasing their chances of falling back onto the planet, while meta-particles ejected in the anti-orbital direction drift away from the planet without having to pass over its surface.

Meta-particles are more likely to form a tail after being ejected in the anti-stellar direction because on that side of the planet, the radiation pressure and centrifugal forces counteracts the planet's gravity.  Conversely, on the stellar side they act in the same direction as the planet's gravity to accelerate meta-particles back towards the planet.  This model does not account for the possibility of the planet shielding meta-particles from the radiation pressure, however we expect that this would only make the preference slightly less pronounced because it would only affect meta-particles that were ejected almost exactly in the anti-stellar direction.

The preference for small ejection angles from the orbital plane ($\vec{Z}$ component close to zero) is mostly because a larger initial velocity component in the $\vec{Z}$ direction reduces the velocity component in the anti-stellar direction.  This means that meta-particles ejected with a large velocity in the $\vec{Z}$ direction require a larger radiation acceleration to escape the planet.  Furthermore, considering the co-rotating reference frame, a larger initial velocity component in the $\vec{Z}$ direction will result in a smaller Coriolis acceleration that can potentially work with the radiation pressure and centrifugal force to help overcome the planet's gravity. Despite it being more likely that meta-particles that are ejected with a large component of their velocity in the $\vec{Z}$ direction will collide with the planet, those that do not collide with the planet set the maximum height of the tail.

This may have interesting implications for understanding the geophysical processes occurring on the planet. It shows that if the planet were relatively massive, even if the particle ejection mechanism acts uniformly over the entire planet's surface, we would only detect the fraction of the total population that was ejected in the particular direction that can form a tail.

\begin{figure*}[h] 
\centering 
\includegraphics[width=18cm]{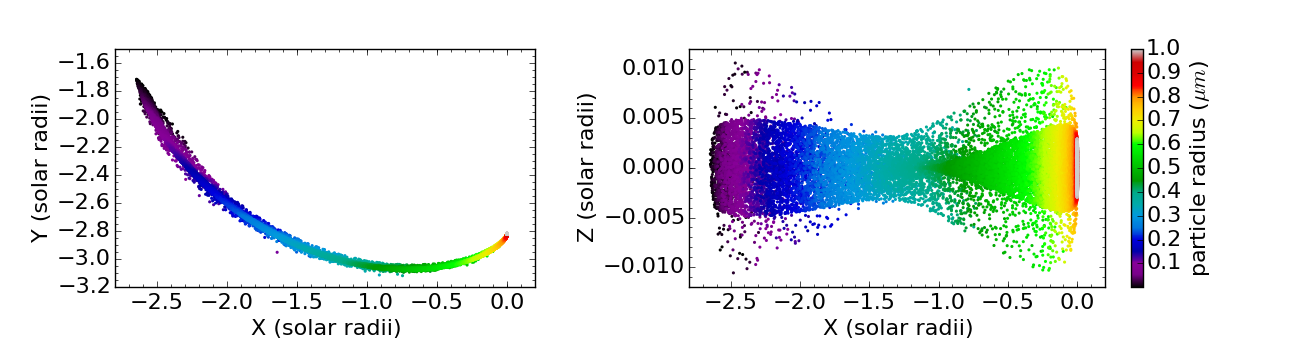} 
\caption{Simulated tail of corundum meta-particles viewed from above the orbital plane (left) and in the orbital plane (right) with meta-particles colour coded according to meta-particle radius.  The left panel's axes have the same scale, however the right panel's vertical axis is stretched  by a factor of $\sim$1000 relative to the horizontal axis because the tail is much longer than it is high.  These meta-particles have an initial radius of 1 $\mu m$ and were ejected from a planet of mass 0.02 M$_\oplus$ in a continuous and spherically symmetric distribution with a velocity of 0.40 times the surface escape (or 1.21 kms$^{-1}$).  The meta-particles were tracked as they sublimated until they were removed when they reached a radius of 1 nm.  The sublimation rate of the meta-particles was set so that they reached a radius of 1 nm after one planetary orbit.  Although there appear to be gaps at high Z positions between individual meta-particles, the grid used for radiative transfer contains at least a few meta-particles per cell and results in a continuous distribution.}
\label{fig:LaptopRun216ParticleSize} 
\end{figure*}

\begin{figure*}[h] 
\centering 
\includegraphics[width=18cm]{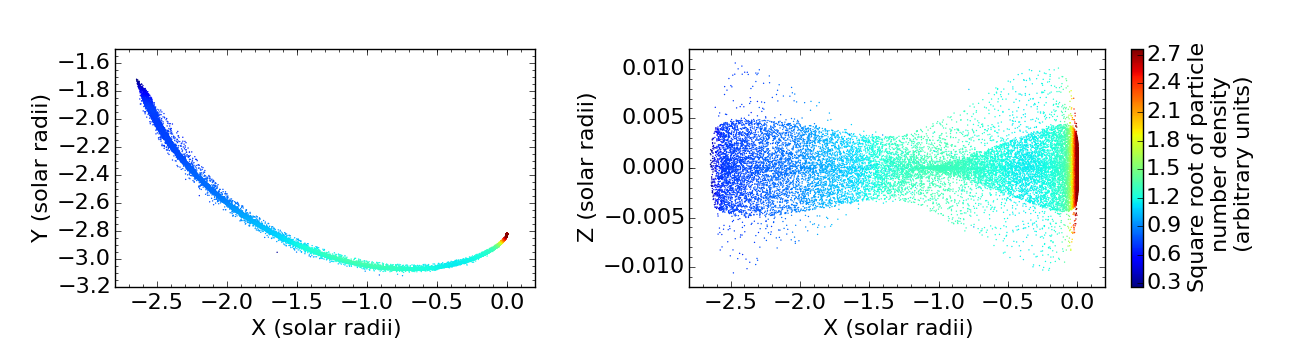} 
\caption{Same as Fig. \ref{fig:LaptopRun216ParticleSize}, but colour coded proportionally to the square root of density to increase the dynamic range.}
\label{fig:LaptopRun216ParticleDensity} 
\end{figure*}

This tail is presented in Figs. \ref{fig:LaptopRun216ParticleSize} and \ref{fig:LaptopRun216ParticleDensity} which show the tail meta-particles colour coded according to meta-particle size and local density.  Despite having a maximum height that is similar to the tail presented in Sect. \ref{sec:OpticallyThin}, this tail has a more rectangular shape, which would diminish the prospect of detecting the double-dip light curve feature (as in Fig. \ref{fig:LC1826}) caused by the dust density enhancement from meta-particles crossing the planet's orbital plane.

\begin{figure*}[h] 
\centering 
\includegraphics[width=16cm]{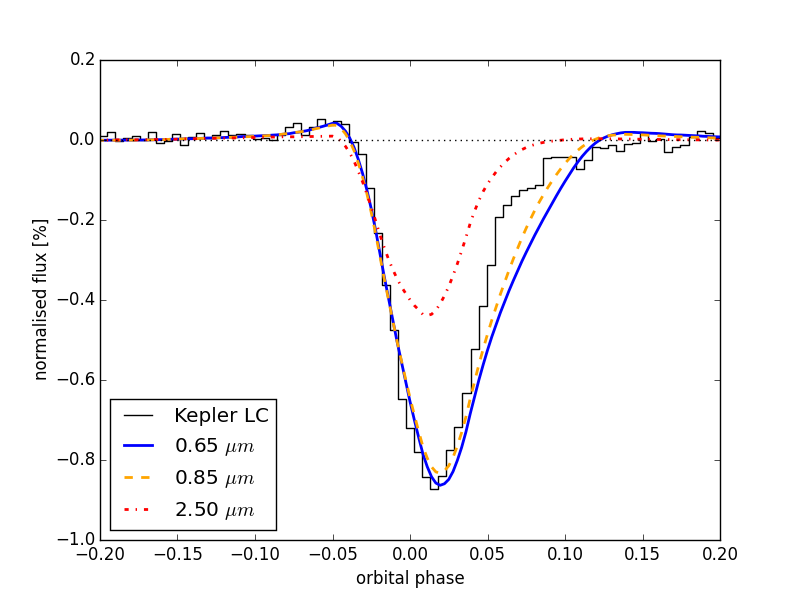} 
\caption{Light curve produced by the tail shown in Figs. \ref{fig:LaptopRun216ParticleSize} and \ref{fig:LaptopRun216ParticleDensity} at wavelengths of 0.65 $\mu$m (solid blue), 0.85 $\mu$m (dashed orange) and 2.5 $\mu$m (dot-dashed red) compared with the Kepler long-cadence light curve of Kepler-1520 b (black).  The model light curves are convolved to the Kepler long-cadence of 30 minutes.  This tail was produced by a planet mass of 0.02 M$_\oplus$ and a meta-particle ejection velocity of 0.40 times the surface escape velocity (or 1.21 km s$^{-1}$).} 
\label{fig:LaptopRun216LC} 
\end{figure*}

The light curve that this tail produces is shown in Fig \ref{fig:LaptopRun216LC}.  To make the simulated light curve have a similar depth to the Kepler average long-cadence light curve for Kepler-1520 b, we scaled the tail dust mass to 3.0$\times10^{14}$ kg, which corresponds to a dust mass-loss rate of 80 M$_\oplus$ Gyr$^{-1}$, only considering the 16\% of meta-particles that actually escape to form a tail.  This implies a lifetime of 0.25 Myr which is also much smaller than the expected lifetimes calculated by \citep{Perez-Becker2013} of 40 $-$ 400 Myr.   

\subsection{Modelling the light curve of Kepler-1520 b with a planet mass of 0.02 M$_\oplus$ and larger maximum height \label{sec:PlanetGrav2}}

\begin{figure*}[h] 
\centering 
\includegraphics[width=18cm]{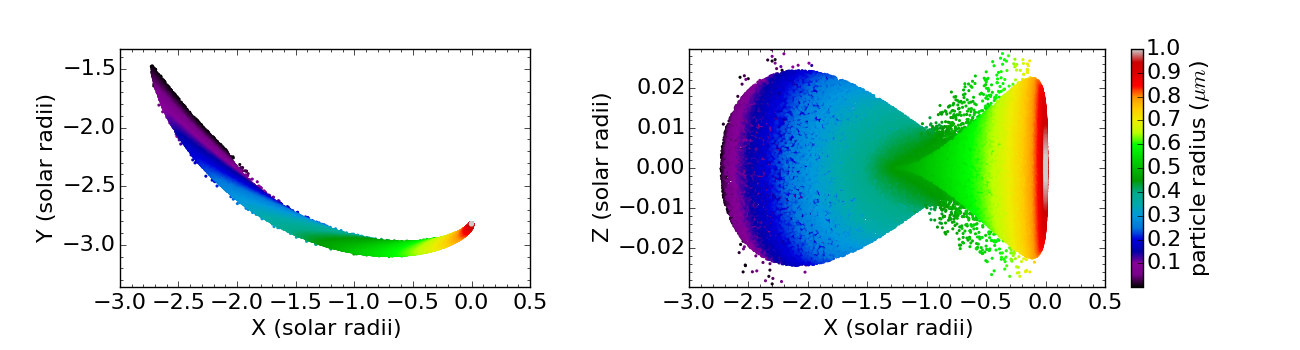} 
\caption{Simulated tail of corundum meta-particles viewed from above the orbital plane (left) and in the orbital plane (right) with meta-particles colour coded according to meta-particle radius.  The left panel's axes have the same scale, however the right panel's vertical axis is stretched  by a factor of $\sim$1000 relative to the horizontal axis because the tail is much longer than it is high.  These meta-particles have an initial radius of 1 $\mu m$ and were ejected from a planet of mass 0.02 M$_\oplus$ and radius 0.28 R$_\oplus$ in a continuous and spherically symmetric distribution with a velocity of 1.034 surface escape velocities (or 3.13 kms$^{-1}$).  The meta-particles were tracked as they sublimated until they were removed when they reached a radius of 1 nm.  The sublimation rate of the meta-particles was set so that they reached a radius of 1 nm after one planetary orbit.} 
\label{fig:RekereRun29ParticleSize} 
\end{figure*}

\begin{figure*}[h] 
\centering 
\includegraphics[width=18cm]{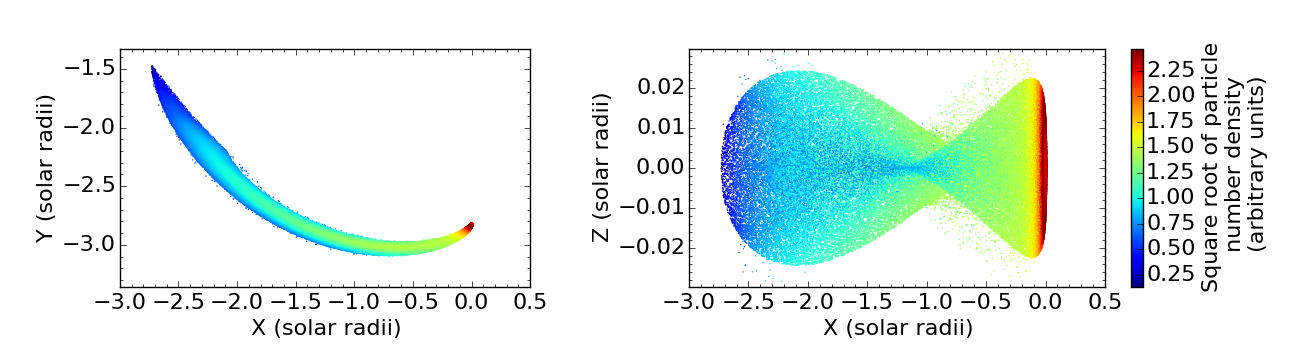} 
\caption{Same as Fig. \ref{fig:RekereRun29ParticleSize}, but colour coded proportionally to the square root of density to increase the dynamic range.} 
\label{fig:RekereRun29Density} 
\end{figure*}

\begin{figure*}[h] 
\centering 
\includegraphics[width=16cm]{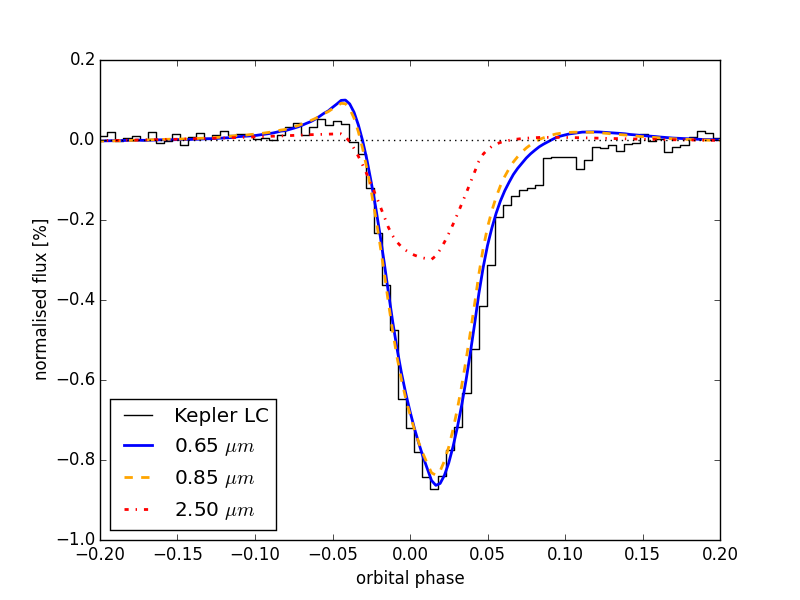} 
\caption{Light curve produced by the tails shown in Figs. \ref{fig:RekereRun29ParticleSize} and \ref{fig:RekereRun29Density}  at wavelengths of 0.65 $\mu$m (solid blue), 0.85 $\mu$m (dashed orange) and 2.5 $\mu$m (dot-dashed red) compared with the Kepler long-cadence light curve of Kepler-1520 b (black).  The model light curves are convolved to the Kepler long-cadence of 30 minutes. This tail was produced by a planet mass of 0.02 M$_\oplus$ and radius 0.28 R$_\oplus$ and an ejection velocity of 1.034 surface escape velocities (or 3.13 kms$^{-1}$).}
\label{fig:RekereRun29Mass1-9211e13LightCurve_65} 
\end{figure*}

For comparison to the simulated tail presented in Section \ref{sec:TailPlanetGrav1} that used a planet mass of 0.02 M$_\oplus$ with a meta-particle ejection velocity of 0.40 times the surface escape velocity, we also simulated a tail with the same planet mass but with the larger meta-particle ejection velocity of 1.034 times the surface escape velocity (or 3.13 kms$^{-1}$).  The simulated tail is presented in Figs \ref{fig:RekereRun29ParticleSize} and \ref{fig:RekereRun29Density} and the resulting light curve is presented in Fig. \ref{fig:RekereRun29Mass1-9211e13LightCurve_65}.  Since the meta-particle ejection velocity is higher than the escape velocity, most of the meta-particles can escape from the planet and form a tail.  Compared to the tail in Section \ref{sec:TailPlanetGrav1}, this tail has more of a bow-tie shape, however it is less well defined than the tail presented in Section \ref{sec:OpticallyThin} due to the planet's larger gravity smearing out the point where the meta-particles' orbital trajectories cross the planet's orbital plane.  After simulating the tail by calculating the meta-particle dynamics (without accounting for radiation shielding through the tail)  we scaled the dust mass of the tail to make it produce the same transit depth as the average long-cadence light curve of Kepler-1520 b of 0.87\%.  The required tail dust mass was 1.92$\times10^{13}$ kg which corresponds to a dust mass-loss rate of 7 M$_\oplus$ Gyr$^{-1}$.  This dust mass-loss rate would result in the planet of mass 0.02 M$_\oplus$ having a lifetime of 2.7 Myr which is more reasonable than the tails presented in the previous sections but still less than the 40 $-$ 400 Myr found by \cite{Perez-Becker2013}.  However, the light curve produced by this more vertically extended tail also over-estimates the pre-ingress forward-scattering peak which prevents us from further decreasing the required dust mass-loss rate by further increasing the tail's height.

\subsection{Behaviour of large particles \label{sec:lareparticles}}

The motion of a dust particle in the tail is controlled by the ratio of the radiation pressure force to the gravitational force, $\beta$ which is a quantity that only depends on radius for a given particle composition and host star spectrum \citep[e.g. Fig. 3 of][]{vanLieshout2014}. In general, $\beta$ becomes very small for large particles of radii  $\gtrsim$10 $\mu$m which results in large particles not being sculpted into a long tail by the radiation pressure.  Therefore, large particles tend to remain around the planet and can drift in front of the planet if they are ejected with some velocity relative to the planet.  

\begin{figure*}[h] 
\centering 
\includegraphics[width=18cm]{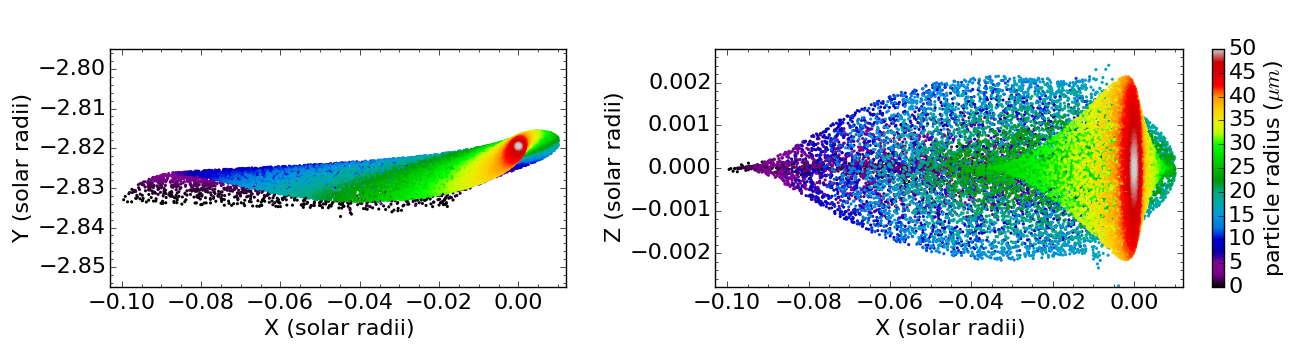} 
\caption{Same as Fig. \ref{fig:1780_radius}, except this tail was simulated with an initial meta-particle size of 50 $\mu$m so the meta-particles do not experience a strong enough radiation pressure to push them into a long tail.} 
\label{fig:LargeParticlesRad} 
\end{figure*}

\begin{figure*}[h] 
\centering 
\includegraphics[width=18cm]{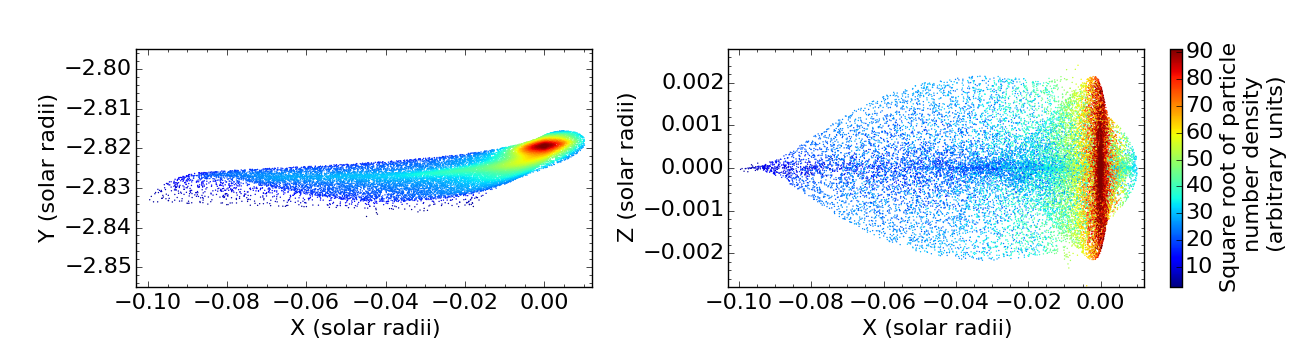} 
\caption{Same as Fig. \ref{fig:LargeParticlesRad}, except the meta-particles are colour coded proportionally to the square root of the density in the tail.} 
\label{fig:LargeParticlesDens} 
\end{figure*}

To illustrate that this can place an upper limit on the allowed particle sizes in the tail, we simulated a tail with an initial meta-particle size of 50 $\mu$m and correspondingly increased the sublimation rate so that the meta-particles completely sublimated after one orbit.  As the large meta-particles sublimate, $\beta$ increases, allowing a small tail to form.  The morphology of this tail is shown in Figs. \ref{fig:LargeParticlesRad} and \ref{fig:LargeParticlesDens}, which show the tail meta-particles colour coded according to the meta-particle radius and square root of meta-particle density, respectively. 

\begin{figure*}[h] 
\centering 
\includegraphics[width=14cm]{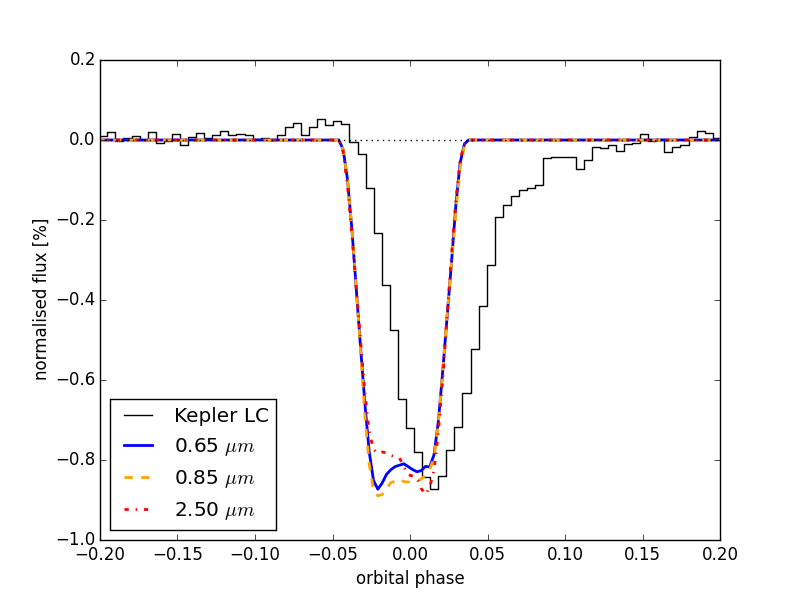} 
\caption{Model light curves for the tail morphology shown in Fig. \ref{fig:LargeParticlesRad} at wavelengths of 0.65$\mu$m (solid blue), 0.85$\mu$m (dashed orange) and 2.5$\mu$m (dot-dashed red) compared with the Kepler long-cadence light curve of Kepler-1520 b (black).  The model light curves are convolved to the Kepler long-cadence of 30 minutes.  To simulate these light curves, the dust meta-particles were ejected with an initial radius of 50 $\mu$m.  Due to 50 $\mathrm{\mu}$m sized meta-particles not experiencing a significant radiation pressure, they do not form a long tail and drift in front of the planet.  This results in a shallow transit depth and early ingress.  To compensate for the reduced absorption resulting from the short tail, these light curves were scaled by a factor of 14 to make them comparable to the Kepler light curve.} 
\label{fig:1820LC} 
\end{figure*}

The transit light curve that this tail produces is shown in Fig. \ref{fig:1820LC}, in which the light curves have been scaled by a factor of 14 to compensate for the reduced cross-section of the shorter tail.  When comparing to the model light curves shown in Fig. \ref{fig:LC1780}, these light curves have an earlier transit time caused by the large number of meta-particles ahead of the planet.  They also have a more symmetric shape due to not having a long tail to produce the gradual increase of flux at egress.  Therefore, this implies from a dust particle dynamics perspective that in order to form a tail long enough to produce an asymmetric transit light curve similar to the Kepler long-cadence light curve of Kepler-1520 b, the particles in the tail must have radii $\lesssim$50 $\mu$m.

%\clearpage

\section{Wavelength dependence \label{sec:WavelengthDependence}}

To investigate how an optically thick tail can influence the wavelength dependence of the transit depth, we calculated the transit depth in several wavelengths as a function of tail dust mass (or mass-loss rate).  We did this by taking the tail configuration presented in Section \ref{sec:OpticallyThin} with a planet mass of 8.35$\times10^{-6}$ M$_\oplus$, as well as a similar tail but with a reduced meta-particle ejection velocity of 272 $ms^{-1}$ (1.2 times the surface escape velocity) and scaling the mass in the tail over three orders of magnitude.  Since we were mainly interested in the transit depth and not the overall shape of the light curve, we saved time by not computing the full light curve to find the transit depth, and instead only carried out the ray-tracing for the viewing angles of phase 0 (mid-transit point) and phase 0.5 to allow the normalised transit depth to be derived.  For the highest tail masses, there is a small signature from the secondary eclipse spanning the orbital phase range of approximately $\phi =$ [0.3,-0.3], which may affect the absolute transit depth of the highest tail masses by $\sim$ $0.01\%$, however the overall trend will be unaffected.   

\begin{figure*}[h] 
\centering 
\includegraphics[width=16cm]{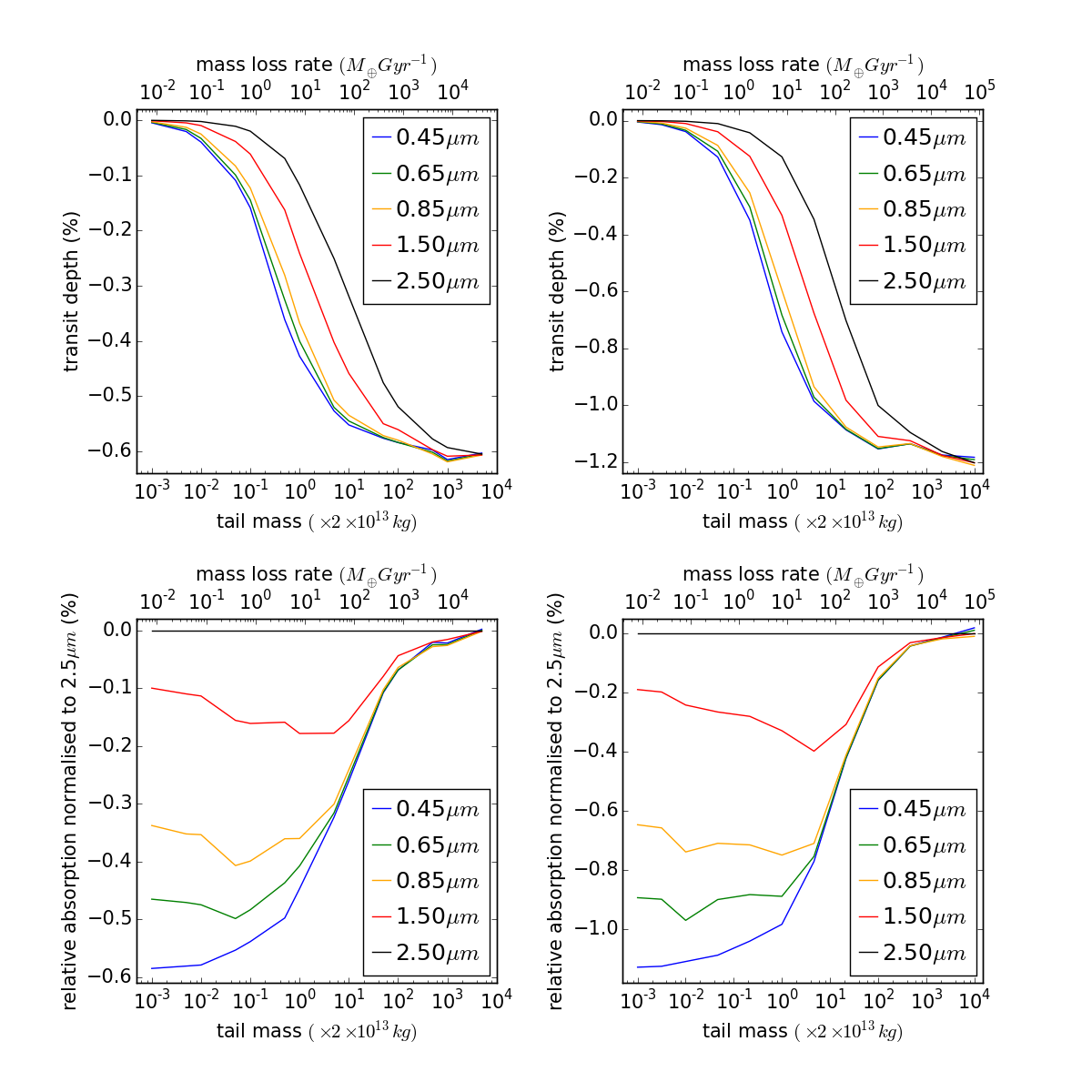} 
\caption{Transit depth in different wavelengths as a function of dust mass in the tail on the bottom horizontal axis or dust mass-loss rate, assuming an initial meta-particle size of 1 $\mu$m and a meta-particle density of 4.02 g cm$^{-3}$, corresponding to the density of corundum, on the top horizontal axis.  The left panels are for a meta-particle  ejection velocity of 272 ms$^{-1}$ and the right panels are for meta-particle ejection velocities of 679 ms$^{-1}$.  The first row shows the absolute transit depth as a function of tail mass. The lower row presents the same data as the upper panels, however all wavelength light curves were normalised to the light curve of 2.5 $\mu$m and were scaled so that every tail dust mass had the same transit depth to highlight the wavelength dependence.} 
\label{fig:WavelengthDependenceCombined} 
\end{figure*}

These results are presented in Fig. \ref{fig:WavelengthDependenceCombined} which comprises four panels.  The left panels are for a meta-particle ejection velocity of 272 ms$^{-1}$ and the right panels are for meta-particle ejection velocities of 679 ms$^{-1}$.  The first row shows the absolute transit depth as a function of tail dust mass and indicates a trend of increasing transit depth with tail dust mass, until the tail becomes optically thick, so that there is no additional absorption from additional mass.  

The lower panels present the same data as the upper panels, however all light curves were normalised to the light curve of 2.5 $\mu$m and were scaled so that every tail dust mass had the same transit depth.  This re-scaling shows that the most wavelength dependence in transit depth occurs for very low-mass tails which are mostly optically thin.  These tails produce very shallow transits, which will be inherently difficult to detect.
Conversely, very high mass tails are optically thick and have almost no wavelength dependence in transit depth.  However, we also predict a range of tail masses from approximately $2\times10^{12}$ $-$ $2\times 10^{14}$ kg that have moderately deep transit depths but still exhibit a significant wavelength dependence.  This suggests the tantalising possibility that if multi-wavelength transit depth observations were to be carried out to an accuracy of about 0.1\%, they could be compared to models such as these and allow another way for the tail dust mass (and mass-loss rate) to be estimated.  However, these results are tailored for Kepler-1520 b under the assumption that its dust composition is corundum.  Therefore, performing this study for a different planet with a different composition and different stellar irradiation may give different results. 

\section{Constraints on particle ejection velocity \label{sec:tailheight}}

As mentioned in Section \ref{sec:doubledip}, the maximum height of an optically thick tail has a large effect on the transit depth. If the tail has sufficient mass to be optically thick, the transit will not depend on the amount of mass in the tail, and instead will only depend on the transiting cross-section of the tail, which is limited by the size of the star and depends on the maximum height and length of the tail.

The length of the tail depends on the lifetime of the particles while the maximum height of the tail depends on the projected height of the tail as seen from the observer, $h$, which is related to the maximum height perpendicular to the orbital plane, $H$, by the orbital inclination, $i$ as $h = H \cos (i)$.  $H$ depends on the component of particle ejection velocity perpendicular to the planet's orbital plane and the mass of the planet due to the planet's gravitational attraction of the ejected particles.  For a spherically symmetric particle outflow from the planet, the tail forms part of a torus with diameter, $H$, which results in $h$ always being equal to $H$ for all viewing inclinations.  However, if the particle outflow were not spherically symmetric, the correcting factor $\cos(i)$ would need to be considered.

Without considering the planet's gravitational attraction, the maximum height of the tail, $H$, can be shown to depend linearly on the vertical component of the particle ejection velocity.  This derivation is given in detail in Appendix \ref{sec:derivation}.  A particle that is ejected from a parent body will also follow a Keplerian orbit that is inclined relative to the orbit of the parent body.  This inclination relative to the parent body's orbit can give a maximum height perpendicular to the orbital plane from trigonometry, which when combined with the inclination formula simplifies to a linear relationship.  The planet's gravity acts to decelerate the ejected particles, but their maximum heights will still depend on their velocity perpendicular to the orbital plane, after deceleration.  This can lead to an apparent non-linear relationship between particle ejection velocity and resulting maximum tail height. 

\subsection{Particle trajectories}

To demonstrate this relationship we simulated tails with a fixed planet mass of 0.02 M$_\oplus$ and ejected meta-particles with a spherically symmetric spatial distribution, while varying the ejection velocity magnitude.  Since we ejected a spherically symmetric stream of meta-particles from the surface of the planet, only meta-particles that have a large component of their velocity perpendicular to the planet's orbital plane attain the maximum height.  However, because of the large number of meta-particles used in these simulations, the tail is optically thick over the entire height of the tail.  In reality, situations could arise where there is an optically thick central band through the tail where it is most dense and optically thin upper and lower edges where it is less dense.    

We calculated the transit depths with MCMax3D as in Section \ref{sec:WavelengthDependence}.  The resulting transit depths and corresponding maximum tail heights are shown in the top and bottom panels of Fig. \ref{fig:PlanetGravTailHeight} respectively.   After simulating the tails without accounting for self-shielding affecting the radiation pressure, we scaled the resulting tail dust masses to the arbitrary large value of 1.2$\times 10^{16}$ kg to ensure that the tail was optically thick so that there would be a constant correspondence between the tail's transit cross-section (set by its maximum height) and transit depth.  However, such a high-mass tail may be unrealistic.  We also examined the maximum tail height and transit depth profile of a mostly optically thin tail of dust mass 2$\times10^{13}$ kg.  This tail produced transit depths that ranged from 0.2 $-$ 1.2\%, but it only approximately had a constant correspondence between transit depth and maximum tail height. The maximum tail height profile has an interesting and non-intuitive shape for meta-particle ejection velocities less than the planet's surface escape velocity because of the interplay between the acceleration terms in Equation \ref{eqn:EOM}.  

For velocities of 1.4 $-$ 2.2 kms$^{-1}$, the meta-particles almost reach the maximum possible height allowed by inclining their orbits, as though the gravitational field of the planet were not present.  This occurs because some meta-particles that are ejected in particular directions (in the co-rotating reference frame) experience sufficient Coriolis and centrifugal accelerations to increase their velocity in the direction perpendicular to the planet's gravitational acceleration enough to allow them to achieve a partial orbit around the planet. The Coriolis and centrifugal accelerations then act during the time of the partial orbit to quickly move these meta-particles radially away from the planet, rapidly decreasing the acceleration due to the planet, and allowing their orbits to incline without having to work against the planet's gravity in the direction perpendicular to the planet's orbit.

For ejection velocities greater than 3.2 kms$^{-1}$, the Coriolis and centrifugal accelerations can not change the increased meta-particle ejection velocities fast enough to allow them to enter partial orbits around the planet.  As a result, they initially work against the gravitational field of the planet until the Coriolis and centrifugal accelerations radially move them beyond the planet's Hill sphere where the acceleration from its gravity is negligible.  The time interval that the planet's gravity is relevant for can be well approximated as the time that a meta-particle with all of its initial velocity perpendicular to the planet's orbital plane (and zero velocity in the radial direction) takes to be accelerated beyond the planet's Hill sphere in the radial direction by the centrifugal acceleration.  The Coriolis acceleration can be neglected for this approximation as it is roughly two orders of magnitude weaker than the centrifugal acceleration.  With an initial centrifugal acceleration of 24 ms$^{-2}$, it takes approximately 700 seconds to cross the planet's Hill sphere.  As a first order approximation, we calculated the resulting vertical velocity of a meta-particle that was decelerated by the planet's surface gravity of approximately 2 ms$^{-2}$ for 700 seconds and found good agreement with the maximum tail height for ejection velocities higher than the surface escape velocity shown in Fig. \ref{fig:PlanetGravTailHeight}.

In the middle region spanning approximately 2.2 $-$ 3.2 kms$^{-1}$, the trajectories of the highest inclination meta-particles transition between the two previously described scenarios.  This involves them being initially decelerated in the radial direction by the planet's gravity enough to allow them to enter a partial orbit around the planet, as was described for the 1.4 $-$ 2.2 kms$^{-1}$ region.  Interestingly, since the meta-particles decelerate until the threshold at which they can enter a partial orbit, this results in the maximum tail height being relatively constant with increasing velocity over this region. This is in contrast to what we see in the region of ejection velocities greater than 3.2 kms$^{-1}$ where the planet's acceleration is not able to reduce the meta-particles' velocities fast enough for them to enter partial orbits.

\subsection{Constraint from the transit depth}

\begin{figure*}
\centering 
\includegraphics[width=12cm]{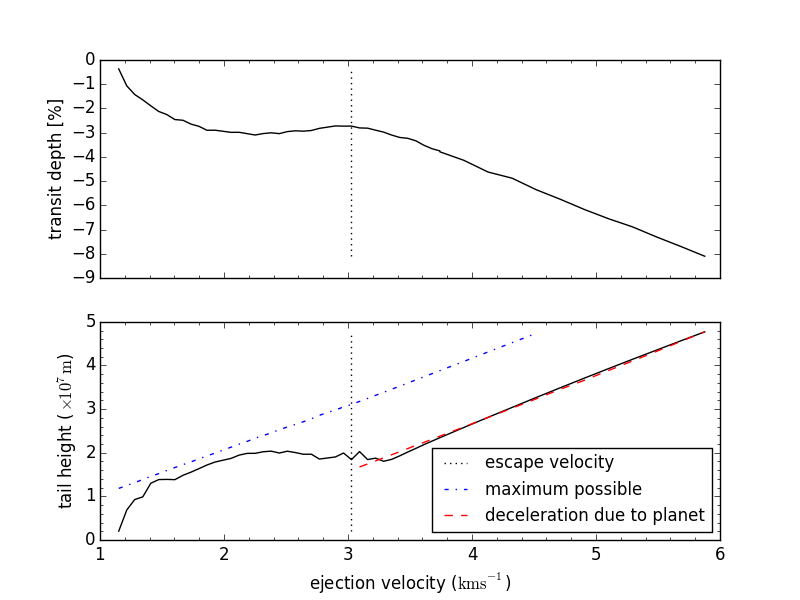} 
\caption{Transit depth (top) and maximum tail height (extent from the planet's orbital plane) (bottom) as a function of meta-particle ejection velocity. The meta-particles were ejected with a spherically symmetric distribution from a planet of mass 0.02 M$_\oplus$ and radius of 0.3 R$_\oplus$.  After the tail was simulated, the dust mass was scaled to the arbitrary large value of 1.2$\times 10^{16}$ kg to ensure that the tail was optically thick. The vertical dotted line is the surface escape velocity.  The blue dash-dotted line is the maximum height that tail meta-particles could attain if they were not decelerated in the vertical direction (perpendicular to the planet's orbital plane) by the planet's gravity.  The dashed red line is the maximum height that tail meta-particles could attain if their vertical velocity were decelerated by the planet's gravity for the time it takes the  meta-particles to radially move beyond the planet's Hill sphere, after being accelerated by the centrifugal acceleration (considering the co-rotating reference frame).}
\label{fig:PlanetGravTailHeight} 
\end{figure*}

The deepest transit depth of Kepler-1520 b as observed by Kepler is approximately 1.4\%.  From Fig. \ref{fig:PlanetGravTailHeight} it can be seen that this transit depth results from an optically thick tail of maximum height from the orbital plane of 1$\times10^{7}$ m, produced by a meta-particle ejection velocity of 1.2 kms$^{-1}$.  Since this is for an optically thick tail that is longer than the stellar diameter, this corresponds to a lower limit on the particle ejection velocity required to produce any given transit depth. The reason for this being a lower limit can be understood by considering an idealised example of a rectangular tail of length $l$ and height $h$ transiting a spherical star of radius $R$.  The transmission through this rectangular tail can be approximated as $T = (1 - f)$ where $f$ is the fractional absorption of the tail, with $f = 1$ representing complete absorption of an optically thick tail and $f < 1$ representing the absorption of an optically thin tail.  

If the tail were optically thick and much longer than the stellar diameter ($f = 1$, $l >> 2R$), the transiting cross-section and hence transit depth will only depend on the projected tail height, which is proportional to the vertical component of the particle ejection velocity.  However, this represents a situation where $l$ and $f$ contribute maximally to the absorption of the tail so if this were not the case and $l$ and $f$ decreased, $h$ would need to increase to compensate for their reduced effect on the total absorption of the tail.  Therefore, the ejection velocity inferred by assuming the tail to be long and optically thick is a lower limit.  The minimum particle ejection velocity of 1.2 kms$^{-1}$ for a planet mass of 0.02 M$_\oplus$ is broadly consistent with the results of \cite{Perez-Becker2013} who found 0.02 M$_\oplus$ to be its most likely mass and typical outflow velocities of $\sim$1 km s$^{-1}$.  However, since \cite{Perez-Becker2013} model a gaseous outflow that gradually accelerates the escaping dust particles, their study is not directly comparable to ours, which ejects meta-particles from the surface of the planet into a vacuum. 

Since the transit depth depends on the tail length, projected tail height (or particle ejection velocity) and optical depth of the dust tail, it will be challenging to disentangle their contributions and determine their individual values.  However, the lower limit on the projected tail height can be used to narrow the allowed parameter space, allowing a more detailed physical interpretation of the tail to be derived.

\subsection{Polarimetry}

Starlight that reflects off disks and planets will become polarised due to being scattered by gas molecules or aerosols.  Therefore, searches for polarimetric signatures can provide valuable information about the structure of disks \citep[e.g.][]{deBoer2017} and cometary coma \citep[e.g.][]{Stinson2016}.  Since the tails of disintegrating rocky exoplanets are composed of small dust particles, they would similarly be expected to induce a polarisation signal.  MCMax3D treats polarisation in its radiative transfer computations so in addition to generating images in non-polarised light, it also generates images in the Stokes $Q$ and $U$ parameters.  This has allowed us to investigate the plausibility of observing the polarisation signal induced by the dust tails of disintegrating rocky exoplanets.   For all of the simulated tails presented in this paper, we examined the normalised polarisation intensity $\sqrt{Q^2+U^2}/I$ (where $I$ is the total intensity) and found that it was generally comparable to the noise from the star, but a weak signal was apparent at the $10^{-5}$ level.  

%\clearpage

\section{Discussion \label{sec:discussion}} 

\subsection{Observational implications}

It is plausible that high-mass tails would be optically thick, while low-mass tails would be optically thin.  This may be a partial explanation for why \cite{Croll2014}, \cite{Felipe2013} and \cite{Schlawin2016} found no evidence for a wavelength dependence in transit depth for transits of comparable depth to the Kepler light curves, while \cite{Bochinski2015} did detect a wavelength dependence in transit depth for similar transit depths.  This scenario would be possible if the material were ejected with variable mass-loss rates and with variable ejection velocities, as is illustrated in Fig. \ref{fig:WavelengthDependenceCombined} which shows, that for a given transit depth, the  tail can be optically thick or thin depending on the maximum tail height.  Therefore, additional multi-wavelength transit observations, including the K band (2.2 $\mu$m) in particular, would be very valuable for better constraining the models.

\subsection{Limitations of the model}

Our model takes about 15 hours to generate a dust tail, tracking 5$\times10^4$ meta-particles and about 80 hours per wavelength to generate a corresponding full phase light curve for that tail model, so it was not feasible to carry out a rigorous parameter space study in an MCMC fashion because the model realisation times are orders of magnitude too long. However, this may be plausible in the future.  Nevertheless, we caution against fitting the average light curve in great detail because of the non-linear relation between the transit light curve and the tail model: a model that explains the average light curve may not correspond to the average of models that would explain the individual transits.

The long light-curve simulation times in our model are in contrast to previously used models (see Introduction), which made approximations to generate transit light curves in a fast way to enable the parameter space to be explored with a MCMC analysis.  This drawback was compensated by MCMax3D offering the advantage of being able to robustly generate transit light curves with part or all of the tail being optically thick.  This enabled us to investigate whether having an optically thick tail could explain why only some  multi-wavelength observations show a wavelength dependence in transit depth.  Although we could not determine a best fit, by assuming reasonable values for most parameters and varying other important parameters in a trial-and-error way, we were able to generate light curves that were a reasonable match to the observed Kepler long-cadence light curve.

All of the simulated tails presented here were produced by ejecting meta-particles with an initial size of 1 $\mu$m, while in reality, particles are probably ejected with a distribution of particle sizes.  This may be related to the discrepancy at egress between the simulated and observed light curve shown in Fig. \ref{fig:LC1780}.  Ejecting meta-particles with a distribution of initial sizes would result in the tail being radially wider because meta-particles of different radii would experience different values of $\beta$ and have different trajectories, as illustrated in Fig. \ref{fig:rosettas}.  However, for meta-particles of corundum in the tail of  Kepler-1520 b, they have a maximum value of $\beta = 0.087$ \citep[Fig. 3. of][]{vanLieshout2014} which is comparable to the range of $\beta$ shown in the right panel of Fig. \ref{fig:rosettas}, indicating that this only has a small effect on the tail's radial width.  Therefore, we do not expect that ejecting meta-particles with a distribution of sizes would significantly change the results of our simulations.  However, this will be further investigated in the forthcoming work, which includes the optical depth in the particle dynamics simulations.

\subsection{High mass-loss rates}

The mass-loss rates that we derive are orders of magnitude higher than those determined by previous studies, which are 0.1 $-$ 1  M$_\oplus$ Gyr$^{-1}$ \citep[e.g.][]{Perez-Becker2013}.  Our higher mass-loss rates are likely related to the optically thick region at the head of the tail (near the planet) that is present in even our mostly optically thin tails.  However, our model neglects extinction caused by gas that could possibly be present after being directly lost from the planet or by being produced by the sublimation of the dust in the tail.  If our model were to include extinction by gas, we would not require as much extinction by dust which would allow a lower dust tail mass or mass-loss rate.  However, estimating the extinction by such gas would rely on many additional assumptions and we simply caution that all of the tail dust masses presented throughout this paper should only be considered as upper limits.

While our derived planet lifetimes appear to be too short to be consistent with the observed occurrence rate of disintegrating rocky exoplanets, this would not be the case if many of these short-lived objects were produced.  However, the occurrence rates for ultra-short period planets derived by \cite{Sanchis-Ojeda2014} suggest that longer lifetimes are required.

\subsection{Constraints from dynamics}
In Section \ref{sec:lareparticles} we show from a dynamical perspective that the particles must be less than approximately 50 $\mu$m to form a tail. While this is based on different physics, it is compatible with the radiative hydrodynamical simulations of \cite{Perez-Becker2013} which showed that large particles are less likely to be present in the tail because they are more difficult to lift out of the planet atmosphere.  It is also consistent with previous observational studies, which all found constraints that varied from $0.1 - 5.6 \mu$m \citep{Brogi2012b,Croll2014, Bochinski2015, vanLieshout2016, Schlawin2016}. 

In addition to the spherically symmetric meta-particle ejection distribution that was used for the tails presented in the preceding sections, we also trialled ejecting meta-particles uniformly from only the day-side and from a 30$^\circ$ cone directed towards the star.  We find that the resulting tails had the same overall morphology as the tails produced by a spherically symmetric distribution, but that there were also some differences.  The day-side only ejection distribution results in a tail that was narrower in the radial direction, while the 30$^\circ$ cone distribution directed towards the star resulted in a tail that was both narrower in the radial direction and the vertical direction (perpendicular to the planet's orbital plane).  Both the reduced radial and vertical extents reduced the extinction from the tail so higher tail masses (or planet mass-loss rates) were required to result in the same transit depths as the tails produced from a spherically symmetric meta-particle ejection distribution.

\subsection{Plausibility of volcanic particle ejection mechanism}

The transit light curves of Kepler-1520 b show that it has transit depths that vary from approximately 0 to 1.4\% and that there is no correlation between consecutive transits \citep{vanWerkhoven2014}.  This could likely be explained by the mass-loss rates varying significantly over time scales comparable to the planet's orbital period. The simulations of \cite{Perez-Becker2013} show that a limit-cycle can plausibly result in the required variation in mass-loss rates on the required time scales. The limit-cycle would form as a result of the mass-loss rate being driven by the stellar flux incident on the planetary surface after passing through the planet's atmosphere of variable opacity.  The maximum mass-loss rate would occur when the atmosphere is clear of dust, which would lead to a dusty atmosphere obscuring the surface and lowering the mass-loss rate until the atmosphere clears.  \cite{Rappaport2012} and \cite{Perez-Becker2013} also qualitatively suggest that the variability may also be partly caused by the limit-cycle being punctuated with unpredictable outbursts from volcanoes or geysers.  However, it is not clear whether this body could sustain sufficient geological activity for this to be a reasonable explanation due to its small size of $<$0.7 R$_\oplus$ \citep{vanWerkhoven2014}. 

The volcanic activity of Io \citep{Lainey2009} and the geyser activity of Enceladus \citep{Hedman2013} both result from tidal interactions with their host planets (Jupiter and Saturn, respectively) and interactions with the other moons in their systems.  Kepler-1520 b is the only known planet in its system so it is unlikely that it will have tidal interactions with other bodies.

Furthermore, the models of \cite{Perez-Becker2013} indicate that bodies with masses $<$0.1 M$_\oplus$ can completely disintegrate in time scales of $\lesssim$10 Gyr, and that Kepler-1520 b is likely in the final few percent of its lifetime so it has probably been at its current small orbital distance for long enough for tidal forces to have circularised its orbit.  A circular orbit is also consistent with the transit timing observations.  Therefore, tidal heating is probably not sufficient to drive any substantial geological activity.  

The simulations of \cite{Perez-Becker2013} showed that small changes in the planet's atmospheric optical depth can lead to large changes in mass-loss rate.  For example, if the optical depth to the surface increased from 0.1 to 0.4, the mass-loss rate would decrease by more than a factor of ten.  Therefore, for a geological process to affect the variability of the transit depth, it would only need to increase the optical depth to the surface by adding more material to the planet's atmosphere, and not need to be energetic enough to eject particles completely from the planet.  If geological activity were to be influencing the transit depth in this way, periods of high geological activity would result in higher optical depths and hence lower mass-loss rates.  This is the opposite scenario to what would be expected for very extreme geological activity directly ejecting particles from the planet, leading to high mass-loss rates during periods of high geological activity.

Typical volcanic eruption velocities on bodies throughout the Solar System are generally consistent with the ejection velocities derived here for Kepler-1520 b.  For example, typical ejection velocities on Earth are of the order of 300 ms$^{-1}$,  while on Mars they are predicted to have been of the order of 500 ms$^{-1}$ and on Venus they are predicted to be about 100 ms$^{-1}$ \citep{Wilson1983}. Furthermore, on Io, due to its high level of geological activity driven by tidal interactions with Jupiter and other moons, the volcanic eruptions have been observed to be 0.5 $-$  1 kms$^{-1}$ \citep{McEwen1983}. However, \cite{Ip1996} found for Io that only dust particles of radii $\leq$$0.01 \mu$m, that are electrically charged, can escape from the volcanic plume and form a dust coma by being accelerated by Jupiter's magnetic field. Therefore, even assuming that a volcanic eruption on Kepler-1520 b was capable of ejecting particles at these high velocities, it is not clear whether it could enable micron sized particles to escape the planet and form a dust tail.

It is not clear from a geological perspective, whether it is reasonable to expect that volcanic outbursts could occur on a body as small as Kepler-1520 b.  However, based solely on the lower limit on particle ejection velocity for Kepler-1520 b that we derived being comparable to the ejection velocities from solar system volcanoes, it could be plausible

\section{Summary \label{sec:summary}}

We have developed a new 3D model of the dust tails of disintegrating rocky exoplanets that ejects sublimating meta-particles from the planet surface to build-up a dust tail, instead of assuming a tail density profile like previous 1D and 2D models.  We generated transit light curves of our simulated tails using the Monte-Carlo radiative transfer code MCMax3D (Min et al. 2009), which accounts for scattering and absorption in a robust way, allowing us to generate transit light curves for optically thick tails. We used this model to investigate how the optical thickness and extent perpendicular to the planet's orbital plane (height) of a general dust tail can affect the observed wavelength dependence and depth of transit light curves.  

We show that there is a decreasing wavelength dependence in transit depth as a function of tail optical depth, potentially explaining why only some multi-wavelength transit observations show a wavelength dependence.  We also find that if the tail is optically thick, the transit depth is not indicative of the amount of mass in the tail, and only depends on the transiting cross-section of the tail.  

Furthermore, we derive that the maximum tail height depends linearly on the vertical (perpendicular to the orbital plane of the planet) component of the particle ejection velocity and derived a lower limit on the particle ejection velocity required to produce a given transit depth.  By applying this to the maximum transit depth of Kepler-1520 b of 1.4\%, we find the required minimum particle ejection velocity to be approximately 1.2 km s$^{-1}$.
We also show from a dynamical perspective that for low ejection velocities, only particles that are ejected in particular directions can escape from the planet and form a tail, and that the particles in the tail must be of radius $\lesssim$50  $\mathrm{\mu}$m. 

To fit the average transit depth of Kepler-1520 b of 0.87\%, we derived dust mass-loss rates of 7 $-$ 80 M$_\oplus$ Gyr$^{-1}$ which are approximately 10 $-$ 100 times larger than those found by previous studies.  Assuming a likely planet mass of 0.02 M$_\oplus$, our dust mass-loss rates imply a planet lifetime up to approximately 3 Myr.  It is unlikely that several objects with such relatively short lifetimes would have been observed so our derived dust mass-loss rates are unrealistic.  The cause of our high dust mass-loss rates is not completely understood but it is likely related to our consideration of optically thick tails that can `hide' mass, while previous studies only considered optically thin tails.  We also did not consider the potential extinction from gas that may be lost from the planet or produced by the sublimation of dust in the tail.  If it were considered, we may require less dust to produce the required extinction, so our high dust mass-loss rates should only be considered as upper limits.
While our large mass-loss rates indicate that more work is required, we believe that these results may help to explain why only some transit observations of Kepler-1520 b show a wavelength dependence and that our constraints on particle ejection velocity give us a more accurate physical interpretation of this intriguing object. 

\appendix{}

\section{Derivation of linear relationship between maximum tail height and vertical velocity \label{sec:derivation}}

If a particle is ejected from a parent body that is on a Keplerian orbit, the ejected particle will follow a Keplerian orbit that is inclined relative to the orbit of the parent body.  This inclination, $i$, is given by \citep[e.g.][]{Fulle1989} 

\begin{equation}\label{eqn:sini}
\sin{\left(i\right)} = \frac{v_z}{\sqrt{v_{\theta}^2+v_z^2}}
,\end{equation}

where $v_z$ is the component of the particle's velocity perpendicular to the orbital plane of the parent body and $v_{\theta}$ is the component of the particle's velocity in the direction of the parent body's orbital velocity.  For particles ejected in the direction of the parent body's orbital angular momentum vector (i.e. out of the parent body's north pole), the particle's ejection velocity is equal to $v_z$ and $v_{\theta}$ is equal to the parent body's orbital velocity.  

If the maximum height of the tail from its lowest to highest particle is $H$, then from trigonometry the maximum height above the orbital plane, $H/2$, that the particles can reach on an orbit with inclination $i$ relative to the parent body's orbital plane is

\begin{equation}\label{eqn:Hover2withi}
\frac{H}{2} = d\tan{\left(i\right)} 
,\end{equation}

where $d$ is the radial distance to the particle, projected onto the ejecting body's orbital plane.  

Substituting equation \ref{eqn:sini} into \ref{eqn:Hover2withi} gives 

\begin{equation}\label{eqn:Hover2}
\frac{H}{2} = d\tan\left(\sin^{-1}\left({\frac{v_z}{\sqrt{v_{\theta}^2+v_z^2}}}\right) \right) 
,\end{equation}

which can be simplified by using the identity 

\begin{equation}
\tan{\left(\sin^{-1}\left(x\right)\right)} = \frac{x}{\sqrt{1-x^{2}}}
\end{equation}

where $x = \frac{v_z}{\sqrt{v_{\theta}^2+v_z^2}}$ to give

\begin{equation}
H = \frac{2d}{v_{\theta}} v_z
,\end{equation}

which shows that $H$ is a linear function of $v_z$. 

Furthermore, if the orbital plane is inclined with an angle $\theta$ relative to the observer, the projected height $h$ will be related to $H$ according to $h = H \cos (\theta)$.  However, if the particle outflow from the planet is spherically symmetric, the tail will approximate part of a torus of diameter, $H$, which will have the same apparent height for all viewing inclinations, $i$.

%-------------------------------------------------------------------
\bibliographystyle{aa} % style aa.bst

\begin{acknowledgements} 
A. R. R.-H. is grateful to the Planetary and Exoplanetary Science (PEPSci) programme of the Netherlands Organisation for Scientific Research (NWO) for support.  I. A. G. S. acknowledges support from an NWO VICI grant (639.043.107).  This work has been supported by the DISCSIM project, grant agreement 341137 funded by the European Research Council under ERC-2013-ADG. We thank the anonymous referees for their constructive comments.
\end{acknowledgements}

\bibliography{disintegrating_planets}

\end{document}